\documentclass[a4paper,11pt]{article}
\usepackage[dvipsnames,svgnames,table]{xcolor}
\usepackage{amssymb}
\usepackage{amsmath}
\usepackage{caption}
\usepackage{mathrsfs}
\usepackage{commath}
\usepackage{comment}
\usepackage{mathtools}

\usepackage{tikz}
\usetikzlibrary{decorations.markings}
\usetikzlibrary{shapes.geometric}
\pgfdeclarelayer{edgelayer}
\pgfdeclarelayer{nodelayer}
\pgfsetlayers{edgelayer,nodelayer,main}

\tikzstyle{double arrow}=[<->]
\tikzstyle{thick}=[-, line width=0.5mm]
\tikzstyle{arrow}=[->]

\tikzstyle{none}=[inner sep=0pt]

\usepackage{url}

\usepackage[margin=3.0cm]{geometry}
\usepackage{graphicx}
\usepackage{hyperref}
\usepackage[numbers,square,sort&compress]{natbib}
\usepackage[binary-units=true]{siunitx}
\usepackage{subcaption}
\usepackage{cleveref}
\crefname{lstlisting}{listing}{listings}
\Crefname{lstlisting}{Listing}{Listings}

\usepackage{csquotes}

\DeclareSIUnit\year{yr}

\usepackage{textcomp}

\usepackage{pgfplotstable,booktabs}

\usepackage[T1]{fontenc}
\usepackage[utf8]{inputenc}
\usepackage{listings}
\usepackage[scaled=0.8]{beramono}

\definecolor{LightGrey}{rgb}{0.9629411,0.9629411,0.9629411}
\definecolor{LighterGrey}{gray}{0.99}
\definecolor{Mauve}{rgb}{0.58,0,0.82}

\newcommand{\ilc}[1]{\mbox{\lstinline[language=Python]{#1}}}

\newcommand{\makeblankentable}[1]{
	\pgfplotstabletypeset[
	font=\tiny,
    columns/0/.style={column name={$\mathfrak{m}$}}, 
    columns/1/.style={column name={$\mathfrak{n}$}},
    columns/2/.style={
        column name=$\mathrm{Nu}$,
        dec sep align,
        /pgf/number format/fixed zerofill,
        /pgf/number format/precision=5
    },
    columns/3/.style={
        column name=$\xi_1$,
        dec sep align,
        /pgf/number format/fixed zerofill,
        /pgf/number format/precision=5
    },
    columns/4/.style={
        column name=$\xi_2$,
        dec sep align,
        /pgf/number format/fixed zerofill,
        /pgf/number format/precision=5
    },
    columns/5/.style={
        column name=$\xi_3$,
        dec sep align,
        /pgf/number format/fixed zerofill,
        /pgf/number format/precision=5
    },
    columns/6/.style={
        column name=$\xi_4$,
        dec sep align,
        /pgf/number format/fixed zerofill,
        /pgf/number format/precision=5
    },
    every head row/.style={
        after row=\midrule
    }
    ]{#1}
}

\pgfplotstableread[col sep = comma]{experiments/tosi/data/blankenbach_case_1_strong.csv}\blankstrongone
\pgfplotstableread[col sep = comma]{experiments/tosi/data/blankenbach_case_2_strong.csv}\blankstrongtwo
\pgfplotstableread[col sep = comma]{experiments/tosi/data/blankenbach_case_3_strong.csv}\blankstrongthree
\pgfplotstableread[col sep = comma]{experiments/tosi/data/blankenbach_case_4_strong.csv}\blankstrongfour
\pgfplotstableread[col sep = comma]{experiments/tosi/data/blankenbach_case_5_strong.csv}\blankstrongfive

\pgfplotstableread[col sep = comma]{experiments/tosi/data/blankenbach_case_1_weak.csv}\blankweakone
\pgfplotstableread[col sep = comma]{experiments/tosi/data/blankenbach_case_2_weak.csv}\blankweaktwo
\pgfplotstableread[col sep = comma]{experiments/tosi/data/blankenbach_case_3_weak.csv}\blankweakthree
\pgfplotstableread[col sep = comma]{experiments/tosi/data/blankenbach_case_4_weak.csv}\blankweakfour
\pgfplotstableread[col sep = comma]{experiments/tosi/data/blankenbach_case_5_weak.csv}\blankweakfive

\newcommand{\maketositable}[1]{
	\pgfplotstabletypeset[
	font=\tiny,
    columns/0/.style={column name={$\mathfrak{m}$}}, 
    columns/1/.style={column name={$\mathfrak{n}$}},
    columns/2/.style={
        column name=$\mathrm{Nu}_\text{bottom}$,
        dec sep align,
        /pgf/number format/fixed zerofill,
        /pgf/number format/precision=5
    },
    columns/3/.style={
        column name=$\mathrm{Nu}_\text{top}$,
        dec sep align,
        /pgf/number format/fixed zerofill,
        /pgf/number format/precision=5
    },
    columns/4/.style={
        column name=$\langle T \rangle$,
        dec sep align,
        /pgf/number format/fixed zerofill,
        /pgf/number format/precision=5
    },
    columns/5/.style={
        column name=$u_\text{rms}$,
        dec sep align,
        /pgf/number format/fixed zerofill,
        /pgf/number format/precision=5
    },
    columns/6/.style={
        column name=$u_\text{surf rms}$,
        dec sep align,
        /pgf/number format/fixed zerofill,
        /pgf/number format/precision=5
    },
    columns/7/.style={
        column name=$\langle \Phi \rangle$,
        dec sep align,
        /pgf/number format/fixed zerofill,
        /pgf/number format/precision=5
    },
    columns/8/.style={
        column name=$\langle W \rangle$,
        dec sep align,
        /pgf/number format/fixed zerofill,
        /pgf/number format/precision=5
    },
    every head row/.style={
        after row=\midrule
    }
    ]{#1}
}

\pgfplotstableread[col sep = comma]{experiments/tosi/data/tosi_case_1_strong.csv}\tosistrongone
\pgfplotstableread[col sep = comma]{experiments/tosi/data/tosi_case_2_strong.csv}\tosistrongtwo
\pgfplotstableread[col sep = comma]{experiments/tosi/data/tosi_case_3_strong.csv}\tosistrongthree
\pgfplotstableread[col sep = comma]{experiments/tosi/data/tosi_case_4_strong.csv}\tosistrongfour

\pgfplotstableread[col sep = comma]{experiments/tosi/data/tosi_case_1_weak.csv}\tosiweakone
\pgfplotstableread[col sep = comma]{experiments/tosi/data/tosi_case_2_weak.csv}\tosiweaktwo
\pgfplotstableread[col sep = comma]{experiments/tosi/data/tosi_case_3_weak.csv}\tosiweakthree
\pgfplotstableread[col sep = comma]{experiments/tosi/data/tosi_case_4_weak.csv}\tosiweakfour

\begin{document}
\title{\bf Automatic weak imposition of free slip boundary conditions via
Nitsche's method: application to nonlinear problems in geodynamics}
\author{
Nathan Sime\thanks{Department of Terrestrial Magnetism, Carnegie Institution for Science (\url{nsime@carnegiescience.edu})}
\and 
Cian R. Wilson\thanks{Department of Terrestrial Magnetism, Carnegie Institution for Science (\url{cwilson@carnegiescience.edu})}
}
\date{}

\maketitle
\begin{abstract}
Imposition of free slip boundary conditions in science and engineering
simulations presents a challenge when the simulation domain is non-trivial.
Inspired by recent progress in symbolic computation of discontinuous Galerkin
finite element methods, we present a symmetric interior penalty form of
Nitsche's method to weakly impose these slip boundary conditions and present
examples of its use in the Stokes subsystem motivated by problems in
geodynamics. We compare numerical results with well established benchmark
problems. We also examine performance of the method with iterative solvers.
\\*[2ex]
\end{abstract}

\section{Introduction}
\label{sec:introduction}

This work is born of a common question asked of the authors. Regarding
engineering and geodynamics finite element analysis, how does one implement
free slip boundary conditions? In particular, the case that the geometry is a
non-trivial shape such as that representative of a `real--world problem'.
Typically the author's recommendation is to use Nitsche's method. However,
this recommendation is met with trepidation regarding the difficulty of
implementation in addition to a reputation for not performing well with
iterative solvers.

The intent of this work to dispel these concerns. Specifically by providing
the mathematical and open source user-friendly software tools to compute
finite element problems using Nitsche's method for (but not limited to)
incompressible flow simulations. Additionally by providing numerical examples
which demonstrate the viability of Nitsche's method for the imposition of free
slip boundary data in geodynamics problems.

\subsection{Nitsche's method}

Nitsche's original work~\cite{nitsche1971} proposed a method for the weak
imposition of Dirichlet boundary conditions on the exterior of a computational
domain. The term `weak imposition' implies that the Dirichlet boundary
condition is not applied as a modification of the underlying finite element
linear algebra system (strong imposition), but rather as a component of the
variational formulation. These additional terms in the variational formulation
comprise exterior facet integrals penalising the difference between the
(unknown) finite element solution and the boundary condition data, and
enforcing consistency and coercivity according to the underlying weak
formulation.

Enforcing boundary conditions in this variational setting permits additional
flexibility. For example, Nitsche's method could be used to enforce
inequalities on the boundary such as that required by the Signorini
problem~\cite{burman2017,wriggers2008formulation}. Furthermore, Nitsche's
method is exploitable in pseudo arc--length continuation of boundary condition
parameters given that the underlying finite element matrix need not be
manipulated~\cite{allgower2012numerical}. Recently Nitsche's method has found
popularity for problems with complex geometries by immersion in a background
mesh using the CutFEM technique~\cite{burman2015}. In this paper we exploit
this flexibility for the imposition of free slip boundary data. Specifically,
given a computational domain, the free slip boundary condition requires that
the component of the unknown vector valued solution which acts parallel to the
domain boundary's normal vector must vanish.

\subsection{Strong and weak imposition of the free slip condition}

\subsubsection{Strong imposition}

The strong imposition of free slip boundary data is trivial when the exterior
boundary of the computational domain aligns with the coordinate system. In
this case, one of the orthogonal components of the vector valued solution may be
strongly enforced to vanish, while the other components are solved in a
natural sense. 

In the case of non-trivial geometries strong imposition is no longer so
straightforward. The underlying finite element stiffness matrix, $\mathrm{A}$,
must be manipulated such that degrees of freedom associated with the vector
valued solution are constrained in the boundary normal direction. For example,
consider the unknown solution vector $\mathrm{u}$, constraint matrix
$\mathrm{C}$ and load vector $\mathrm{b}$, the finite element system becomes:
find $\mathrm{u}$ such that
\begin{equation}
\mathrm{C^\top A C u} = \mathrm{C^\top b}, \label{eq:constraint_matrix}
\end{equation}
where the prefactor $\mathrm{C}^\top$ preserves symmetry. The matrix product
operation on the left hand side of \cref{eq:constraint_matrix} is
computationally expensive. Therefore the operator $\widehat{\mathrm{A}} =
\mathrm{C^\top A C}$ should be assembled directly when constructing the finite
element system.

\begin{figure}
\centering
\begin{subfigure}{.75\textwidth}
  \centering
  \includegraphics[width=1.\linewidth]{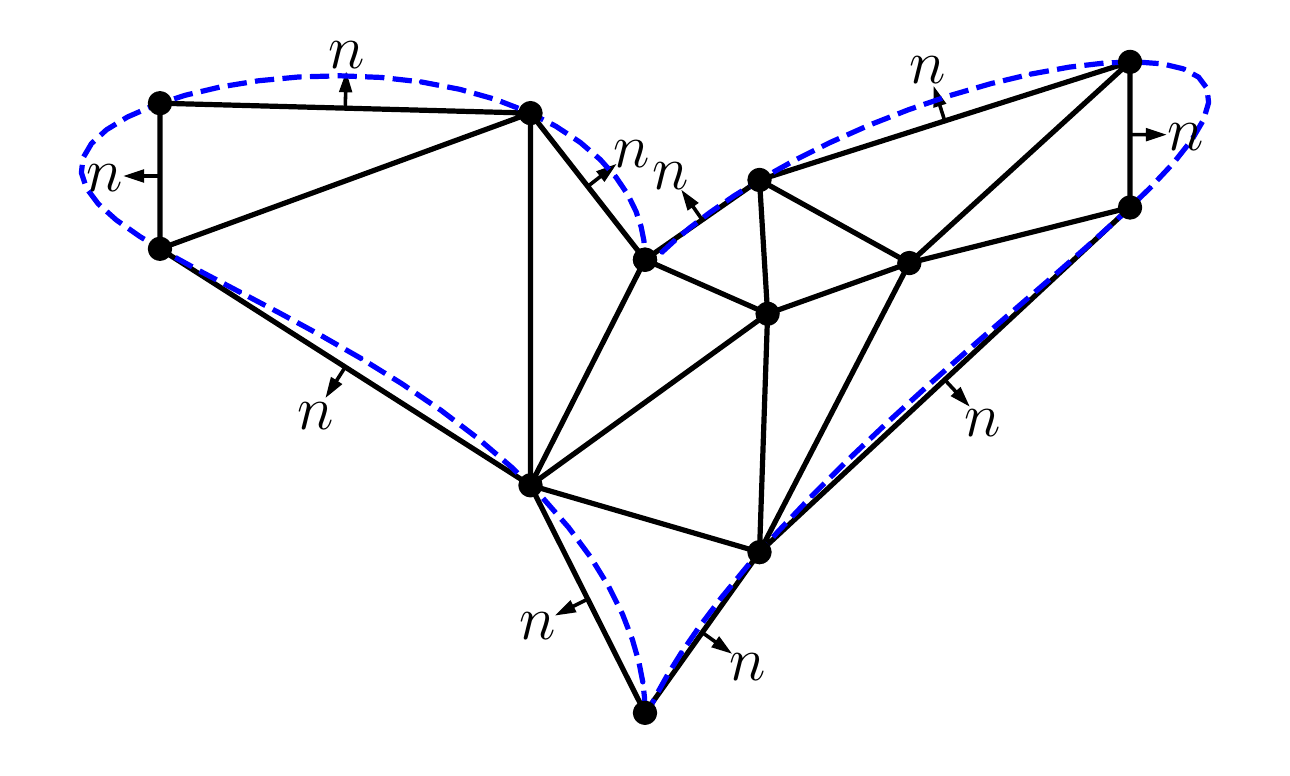}
\end{subfigure}
\caption{Example mesh discretisation using a piecewise linear boundary
representation. The true boundary is indicated by the dashed line. $n$
denotes the outward pointing normal unit vector on the mesh exterior. The
degrees of freedom of a standard piecewise linear finite element function
space are defined at the vertices of the mesh. Consider how the facet normal
is not defined at these locations which is problematic for strong imposition
of free slip boundary conditions.}
\label{fig:meshing_issue}
\end{figure}

When imposing free slip boundaries in a strong sense, we must also take into
account that the outward pointing unit normal vector on the boundary is not
well defined for all finite elements. Consider a vector valued solution sought
in a standard piecewise linear finite element vector space. The degrees of
freedom are defined at the vertices of the mesh. At these degrees of freedom
the facet normal is not well defined, see \cref{fig:meshing_issue} for
example. Special treatment of these cases could be considered by, for example,
taking the average of the two neighbouring facets. However, what approach should
be taken at the intersection of three or more boundaries? This requirement
presents itself in a numerical example shown later in
\cref{sec:subduction_zone}.

The implementation challenge for strong imposition lies in the efficient
assembly of $\widehat{\mathrm{A}}$, the clear definition of the facet normal,
and a user--friendly interface for the definition of boundary conditions
generating~$\mathrm{C}$, cf.~\cite{bangerth2009data}.

\subsubsection{Lagrange multiplier}

Another approach to the imposition of free slip boundary data is by Langrange
multiplier. In this setting a new Langrage multiplier variable is introduced
to enforce the free slip boundary data as a system constraint in the
variational formulation. This mathematical setting provides a flexible
framework for the imposition of boundary data, however, the expanded
variational formulation causes the underlying linear system to the grow by an
additional block row and column. Additionally this block must be carefully
considered when preconditioning the underlying linear system. We refer
to~\cite{urquiza2014} for a summary of the use of Lagrange multiplier methods
for free slip boundary data in the Stokes system.

\subsubsection{Nitsche's method}

Nitsche's method is attractive for free slip boundary conditions. The normal
and tangential components of the vector field on each boundary can be
separated and prescribed in a mathematically consistent
setting~\cite{nitsche1971,urquiza2014,unifieddg}. Specifically without the
need for definition of a facet normal at mesh vertices. There is no
requirement for the constraint matrix~$\mathrm{C}$ to be constructed. Simply
assembly of the stiffness matrix~$\mathrm{A}$. And there is no requirement for
growth in the block size of the underlying linear system, also simplifying
preconditioning. The implementation challenge lies in a user--friendly
interface for defining a boundary condition by Nitsche's method. We address
this in \cref{sec:implementation}.

\subsection{Geodynamics models}

The conservation equations underlying geodynamics models typically
encapsulate the Stokes system. In turn, these models require free slip
boundary conditions, for example at the core-mantle-boundary and
lithosphere~\cite{blankenbach1989,tosi2015}, or on a dipping slab in a
subduction zone~\cite{vankeken2008}.

Prior works demonstrate the use of Nitsche's method for the imposition of free
slip velocities in the Stokes system in a linear
setting~\cite{urquiza2014,Massing2014,hansbo2002unfitted,HANSBO2014}. However,
modern geodynamics simulations are concerned with highly nonlinear temperature
and strain rate dependent viscosity models.

At this point, we highlight that the discontinuous Galerkin (DG) method can be
considered a generalisation of Nitsche's method. Drawing our attention to the
DG literature, we find that imposition of free slip boundary conditions in
nonlinear compressible flow problems is commonplace, for
example~\cite{Hartmann2002}. Therefore we seek to marry the presentation of
Nitsche's method for the Stokes system with the works which examine the
formulations of DG finite element method (FEM) for nonlinear problems.

Whilst doing this, we must ensure that the extra terms arising from Nitsche's
method will be formulated automatically and in a familiar and `friendly'
manner in the computational framework. Ultimately, the utility of Nitsche's
method in esoteric modelling is strongly dependent on its ease of use.

\subsection{Implementation}

A key detriment of Nitsche's method is its difficulty to implement. In the
case of a problem such as the finite element discretisation of Poisson's
equation, the additional facet integral terms are straightforward to add to a
finite element code. However, for more complex systems such as
hyperelasticity, incompressible and compressible flow and electromagnetics,
the additional facet terms are verbose and prone to human error. Introducing
nonlinear material coefficients (such as diffusivity or viscosity) yields a
finite element formulation arduous to write on paper, let alone in code.

The challenge therefore presents itself. We seek to homogenise the
mathematical formulation of Nitsche's method for these classes of finite
element problems. Thereby we construct a computational tool for automatically
applying boundary conditions by Nitsche's method, agnostic of the underlying
finite element problem.

The former issue of homogenisation of the mathematical problem is demonstrated
in works such as~\cite{Hartmann2002,hartmann2007adjoint,nate2018}. We repeat
these formulations later in this work.
Computational symbolic representation of the underlying finite element
formulation is used to address the latter issue for constructing a
computational tool. In this setting the finite element formulation is
represented in computer code which is intended to be high-level, familiar and
recognisible from the mathematical formulation. From this representation high
performance code for the assembly of the finite element matrix and vector is
automatically generated. In this vein, a particularly popular open source
framework is the Unified Form Language~(UFL)~\cite{ufl}. For example, UFL has
been adopted by the following open source finite element suites: FEniCS and
FEniCS--X (used in this work)~\cite{AlnaesBlechta2015a,fenicsx}, the Firedrake
project~\cite{firedrake}, DUNE-FEM~\cite{dunefem} and
TerraFERMA~\cite{terraferma}. Other projects have developed their own
framework for automatic code generation, for example FreeFem++~\cite{freefem},
NGSolve~\cite{ngsolve} and AptoPy/FEM~\cite{nate_phd}.

Despite the advantages of the computational symbolic algebra representation,
the implementation of Nitsche exterior facet terms is still verbose and prone
to human error. Houston and Sime's prior work with automatic symbolic
formulation of DG methods addresses this issue in the development of the
\ilc{dolfin\_dg} library~\cite{nate2018}. They demonstrate how finite
element packages with computational symbolic representation of the variational
formulation can be exploited to add another layer of abstraction by
automatically generating the DG finite element formulation itself.

The paradigm of automatic finite element \emph{formulation--formulation} will
be exploited here. The strong similarity between the weak imposition of
exterior Dirichlet boundary conditions by the DG finite element method and
Nitsche's method allow us to use the designs exhibited in~\cite{nate2018}.

\subsection{Iterative solvers}

Although the automatic formulation of the Nitsche terms in the weak
formulation is useful, the generated code for assembling the finite element
system must perform well. For example, many modern geodynamics models require
discretisations of equations defined in complex 3D geometries. Meshes of these
geometries must be of sufficient granularity and fidelity which give rise to
very large numbers of unknowns.

The FEniCS and FEniCS--X finite element software is used in this work. It has
been demonstrated to be scalable when solving large systems where the degree
of freedom count exceeds~\num{e9}~\cite{richardson:2013,richardson2019}.
Therefore we seek to ensure that the use of automatically generated Nitsche
terms does not impact this scaling.

A key component of scalable finite element analysis is the use of iterative
solvers. For example Krylov subspace solvers combined with algebraic multigrid
preconditioners. The standard Nitsche term which penalises the difference
between the unknown solution and the boundary data incorporates a penalty
parameter. Indeed, those unfamiliar with Nitsche's method may recognise the
penalty method when ignoring the consistency and coercivity terms. However,
the penalty method's penalty parameter would typically be a large number which
can be detrimental to the condition number of the underlying finite element
matrix. In turn this could impact the utility of iterative methods causing
Krylov methods to stagnate or diverge~\cite{saad2003iterative}. By use of
Nitsche's method, this penalty parameter may be orders of magnitude smaller,
yielding a (typically) well conditioned linear system ripe for solution by
exploiting iterative methods~\cite{juntunen2009}. This is explored in the
ultimate numerical example of this paper in \cref{sec:scaling}.

\subsection{Article outline}

The flow of this paper is as follows: In \cref{sec:sipg_dirichlet} we form a
foundation of Nitsche's method for Dirichlet boundary data in a general sense
for elliptical partial differential equation (PDE) problems. The specialised
case for Dirichlet boundary data in a direction perpendicular to a domain's
boundary is presented in \cref{sec:sipg_freeslip}. With this foundation we
formulate the Boussinesq approximation in \cref{sec:problem_definition} in
addition to free slip boundary conditions and thereby its finite element
formulation. We highlight the benefits of the automatic formulation of the
Nitsche boundary terms using computational symbolic algebra by the UFL and
\ilc{dolfin\_dg} libraries in \cref{sec:implementation}. With all the tools
in place, we demonstrate numerical experiments in \cref{sec:experiments}.
Finally we provide some concluding remarks in \cref{sec:conclusion}.

\section{Weak imposition of the slip boundary condition}
\label{sec:weak_imposition}

In this section we examine the formulation of Nitsche's method in a general
setting for nonlinear elliptic problems. Initially in \cref{sec:definitions}
we define the general nonlinear elliptic equation. Inspired by the DG FEM
literature, a symmetric interior penalty Galerkin (SIPG) approach will be
taken~\cite{unifieddg} while noting here that we are not limited to solely the
SIPG formulation. In~\cref{sec:sipg_dirichlet} the general formulation of the
weak imposition of Dirichlet data on the exterior boundary is presented as
summarised in prior work~\cite{nate2018}. Using this as a foundation, we then
focus on the case that normal and tangential data should be prescribed
separately in~\cref{sec:sipg_freeslip}. This yields the semilinear
formulation for imposition of the free slip boundary condition.

\subsection{Elliptic nonlinear boundary value problem}
\label{sec:definitions}

Let $\Omega \subset \mathbb{R}^d$, $d=2, 3$, be the domain of interest,
e.g.~the Earth's mantle. The boundary of the domain is $\partial\Omega =
\Gamma$ which is subdivided into Dirichlet, Neumann and `slip'
components, $\Gamma_D$, $\Gamma_N$ and $\Gamma_S$, respectively. By `slip'
boundary we mean that the normal and tangential components are prescribed in
the Dirichlet and Neumann sense, respectively, on $\Gamma_S$. The boundary
components do not overlap $\Gamma_D \cap \Gamma_S \cap \Gamma_N = \emptyset$
and the Dirichlet boundary must either be non-empty $\Gamma_D \neq \emptyset$
or the tangent of the `slip' boundary must span at a minimum $d$ non--parallel
planes. Let $n$ be unit vector pointing outward and normal to the boundary
$\Gamma$.

The abstract nonlinear elliptic PDE of interest is to find $u$ such that:
\begin{equation}
- \nabla \cdot \mathcal{F}^v(u, \nabla u) = f \quad \text{ in } \Omega \label{eq:gen_visc_eq}
\end{equation}
complemented by boundary conditions
\begin{align}
u &= u_D  &\text{ on } \Gamma_D, \\
\mathcal{F}^v(u, \nabla u) \cdot n &= g_N  &\text{ on } \Gamma_N, \\
u \cdot n &= u_S \cdot n &\text{ on } \Gamma_S, \\
(\mathcal{F}^v(u, \nabla u) \cdot n) \cdot \tau_i &= g_{\tau,i}, \quad i=1,\ldots,d-1, &\text{ on } \Gamma_S,
\end{align}
where $u_D$, $g_N$, $u_S$ and $g_{\tau,i}$, $i=1,\ldots,d-1$, are Dirichlet,
Neumann, normal Dirichlet and tangential Neumann data, respectively. $\tau_i$,
$i=1,\ldots,d-1$, are the orthonormal vectors which span the plane tangential
to $\Gamma_S$.

\subsection{Finite element formulation}

The domain $\Omega$ is subdivided into non-overlapping elements $\kappa$ of
size $h_\kappa$. These elements comprise a mesh $\mathcal{T}^h = \cup_{\kappa}
\kappa$. Let $V^{h,\ell}$ be the finite element space of piecewise polynomials
defined on each element $\kappa \in \mathcal{T}^h$ of degree $\ell \in
\mathbb{N}$ and continuous in $\Omega$. We use the notation $(u, v) =
\int_\Omega u \cdot v \dif{x}$ to define the inner product in $\Omega$, and on
the exterior boundary $(u, v)_\Gamma = \int_{\Gamma} u \cdot v \dif{s}$.

The finite element formulation of the system~\cref{eq:gen_visc_eq} is to find
$u_h \in V^{h,\ell}$ such that
\begin{equation}
\mathcal{N}(u_h, v) \coloneqq 
\mathcal{N}_\Omega(u_h, v) 
+ \mathcal{N}_{\Gamma_D}(u_h, v) 
+ \mathcal{N}_{\Gamma_N}(u_h, v)
+ \mathcal{N}_{\Gamma_S}(u_h, v) \equiv 0
\quad \forall v \in V^{h,\ell}. \label{eq:total_residual}
\end{equation}
Here the semilinear residual terms $\mathcal{N}_\Omega(\cdot, \cdot),
\mathcal{N}_{\Gamma_D}(\cdot, \cdot), \mathcal{N}_{\Gamma_N}(\cdot, \cdot)$
and $\mathcal{N}_{\Gamma_S}(\cdot, \cdot)$ correspond to the volume, Nitsche
Dirichlet, Neumann and Nitsche `slip' boundary terms, respectively. Clearly at
this stage we are able to define
\begin{align}
\mathcal{N}_\Omega(u, v) &\coloneqq (\mathcal{F}^v(u, \nabla u), \nabla v) - (f, v), \label{eq:fem_volume_terms}\\
\mathcal{N}_{\Gamma_N}(u, v) &\coloneqq - (g_N, v)_{\Gamma_N}. \label{eq:neumann_terms}
\end{align}
The remaining terms, $\mathcal{N}_{\Gamma_D}(\cdot, \cdot)$ and
$\mathcal{N}_{\Gamma_S}(\cdot, \cdot)$ will be defined in the following
sections.

\subsection{Symmetric interior penalty Galerkin method for Dirichlet boundary
data}
\label{sec:sipg_dirichlet}

In this section we will define~$\mathcal{N}_{\Gamma_D}(\cdot, \cdot)$. We
recall the SIPG method for the weak imposition of Dirichlet boundary data and
refer to the review~\cite{unifieddg} and the monographs cited therein for
details and analysis. For more details of the formulation presented in this
section we refer to~\cite{nate2018}.

Firstly we homogenise~\cref{eq:gen_visc_eq} such that 
\begin{equation}
-\nabla \cdot \mathcal{F}^v(u, \nabla u)  = -\nabla \cdot \del{G \nabla u},
\label{eq:gen_visc_eq_homo}
\end{equation}
where $G$ is the homogenity tensor defined by
\begin{equation}
G_{kl} \coloneqq \frac{\partial \mathfrak{f}^v_k}{\partial \del{\nabla u}_l}, \quad k, l = 1,\ldots,d,
\end{equation}
which is written in terms of the columns of the viscous flux operator
\begin{equation}
\mathcal{F}^v(u, \nabla u)  = (\mathfrak{f}^v_1(u, \nabla u),
\ldots, \mathfrak{f}^v_d(u, \nabla u))
\end{equation}
and we use the hyper-tensor product (in terms of Einstein summation notation)
\begin{equation}
(G \nabla u)_{ik} = (G_{kl})_{ij} (\nabla u)_{jl}.
\end{equation}
Here, we are concerned only with applying the weak imposition of boundary data
$u_\Gamma(u) = u_D$ on the Dirichlet boundary $\Gamma_D$. The Nitsche boundary
formulation in terms of the homogeneity tensor is given by
\begin{align}
\mathcal{N}_{\Gamma_D}(u, v) \coloneqq
&- \del{\mathcal{F}^v(u_\Gamma, \nabla u), v \otimes n}_{\Gamma_D} \nonumber\\
&- \del{(u - u_\Gamma) \otimes n, G^\top_\Gamma \nabla v}_{\Gamma_D} \nonumber\\
&- C_\mathrm{IP} \frac{\ell^2}{h_F} \del{G_\Gamma (u - u_\Gamma) \otimes n, v \otimes n}_{\Gamma_D},
\label{eq:nitsche_form}
\end{align}
where $G_\Gamma = G(u_\Gamma (u))$ and the transpose product $(G^\top
\nabla v)_{jl} = (G_{kl})_{ij} (\nabla v)_{ik}$, $h_F$ is the measure of the
local facet $F$ which discretises $\partial\Omega$ and $C_\mathrm{IP}$ is a
sufficiently large constant interior penalty parameter which is independent of
the mesh chosen to be $C_\mathrm{IP} = 20$ in this work~\cite{houston2003cip}.

\subsection{SIPG-weak imposition of boundary data in the facet normal direction}
\label{sec:sipg_freeslip}

The previous section provides a familiarity for the weak imposition of
Dirichlet boundary data on $\Gamma_D$. Now consider that only the component of
the numerical flux acting perpendicular to the boundary $\Gamma_S$ be
prescribed. This leaves the remaining
tangenital component on $\Gamma_S$ to be naturally enforced.

We define the normal projection and rejection operators acting on a vector
$v \in \mathbb{R}^d$
\begin{align}
P_n (v) &\coloneqq (n \otimes n) \cdot v, \\
P_\tau (v) &\coloneqq v - P_n(v) = (I - n \otimes n) \cdot v,
\end{align}
respectively. One can intuitively interpret $P_n(v)$ and $P_\tau(v)$ as the
normal and tangential components of $v$, respectively. On $\Gamma_S$ the
normal component of the numerical flux $\mathcal{F}^v(u, \nabla
u) \cdot n$ needs to be conserved. This numerical flux is written in terms of
normal and tangential components as follows
\begin{align}
\mathcal{F}^v(u, \nabla u) \cdot n &= P_n(\mathcal{F}^v(u, \nabla u) \cdot n) + P_\tau(\mathcal{F}^v(u, \nabla u) \cdot n) \nonumber \\
&= P_n(\mathcal{F}^v(u, \nabla u) \cdot n) + \sum_{i=1}^{d-1} g_{\tau,i} \tau_i. \label{eq:flux_norm_tanj}
\end{align}
Furthermore we note the following identity
\begin{equation}
(v \otimes n, \mathcal{F}^v(u, \nabla u)) 
= (v_i n_j, \mathcal{F}^v(u, \nabla u)_{ij})
= (v_i, \mathcal{F}^v(u, \nabla u)_{ij} n_j)
= (v, \mathcal{F}^v(u, \nabla u) \cdot n). \label{eq:outer_inner_ident}
\end{equation}
Applying \cref{eq:flux_norm_tanj,eq:outer_inner_ident} to the Nistche boundary
terms in \cref{eq:nitsche_form} we arrive at the following integrals on the
boundary $\Gamma_S$
\begin{align}
\mathcal{N}_{\Gamma_S}(u, v) \coloneqq
&- \sum_{i=1}^{d-1} \del{g_{\tau,i}, v \cdot \tau_i}_{\Gamma_S} \nonumber\\
&- \del{P_n(\mathcal{F}^v(u_\Gamma, \nabla u) \cdot n), v}_{\Gamma_S} \nonumber\\
&- \del{u - u_\Gamma, P_n((G^\top_\Gamma \nabla v) \cdot n) }_{\Gamma_S} \nonumber\\
&- C_\mathrm{IP} \frac{\ell^2}{h_F} \del{P_n((G_\Gamma (u - u_\Gamma)) \cdot n), v}_{\Gamma_S}.
\label{eq:nitsche_slip_form}
\end{align}
On $\Gamma_S$ the boundary flux function is set such that solely the normal
component is prescribed data, i.e., $u_\Gamma(u) = u_S$.

\section{Boussinesq approximation}
\label{sec:problem_definition}

With the general formulation \cref{eq:gen_visc_eq} and its finite element
formulation \cref{eq:total_residual} and terms defined in 
\cref{eq:fem_volume_terms,eq:neumann_terms,eq:nitsche_form,eq:nitsche_slip_form}
we move on to the specific case of the steady state incompressible Boussinesq
approximation. The Boussinesq approximation underpins mantle convection
simulations of buoyancy driven flow. We seek the non--dimensionalised
velocity~$u$, pressure~$p$ and temperature~$T$ such that equations for
conservation of momentum, mass and energy are satisfied
\begin{align}
-\nabla \cdot \left( 2 \eta \varepsilon(u) - p I\right) &= f &\text{ in } \Omega, \label{eq:stokes_vel} \\
\nabla \cdot u &= 0 &\text{ in } \Omega, \label{eq:stokes_cont} \\
- \nabla \cdot \del{\kappa \nabla T} + u \cdot \nabla T &= Q &\text{ in } \Omega. \label{eq:heat}
\end{align}
Here $Q$ is an external heat source, $\varepsilon(u) = \frac{1}{2}(\nabla u +
{\nabla u}^\top)$ is the rate of strain tensor, $I_{ij} = \delta_{ij}$, $i,
j=1,\ldots,d$, is the identity tensor, $f$ is a momentum source and
$\eta$ is the viscosity. We highlight that the viscosity is typically a
nonlinear function of the temperature and velocity, i.e., $\eta = \eta(u, T)$.
This mandates that the system~\cref{eq:stokes_vel,eq:stokes_cont,eq:heat} is
nonlinear.

\subsection{Stokes subsystem finite element formulation}

We focus our attention on the Stokes subsystem
in~\cref{eq:stokes_vel,eq:stokes_cont}. Here $\mathcal{F}^v(u, \nabla u) = (2
\eta \varepsilon(u) - p I)$. Discretising, the finite element formulation
reads: find $(u_h, p_h, T_h) \in W^h$ where $W^h$ is an appropriate mixed
element such that
\begin{align}
\mathcal{N}_\text{Boussinesq}(u_h, p_h; v_h, q) &\vcentcolon= 
\mathcal{N}_\text{momentum}(u_h, p_h, T_h; v) \nonumber \\
&+ \mathcal{N}_\text{mass}(u_h, p_h, T_h; q) \nonumber \\
&+ \mathcal{N}_\text{energy}(u_h, p_h, T_h; s) \equiv 0 \nonumber \\
&\quad \forall (v, q, s) \in W^{h}.
\label{eq:form_boussinesq_total}
\end{align}
The formulations $\mathcal{N}_\text{momentum}$ and $\mathcal{N}_\text{energy}$
are formed from \cref{eq:stokes_vel,eq:heat} in a straightforward manner
according to \cref{eq:total_residual}. The special case of the first order
continuity formulation $\mathcal{N}_\text{mass}$ is given by
\begin{equation}
\mathcal{N}_\text{mass}(u, q) \coloneqq
(\nabla \cdot u, q) - ((u - u_\Gamma) \cdot n, q)_{\Gamma_D \cup \Gamma_S},
\label{eq:cont_fem_form}
\end{equation}
see \cite{cockburn2002local,nate_phd}, for example. Two choices of $W^h$ are
employed in this work:
\begin{itemize}
\item The standard Taylor--Hood element where $W^{h,\text{TH}} =
[V^{h,\ell}]^d \times V^{h,\ell-1} \times V^{h,\mathfrak{s}}$ and $\ell \ge
2$ and $\mathfrak{s} \ge 1$~\cite{taylor1973numerical}.
\item The suboptimal piecewise constant pressure approximation where $W^{h,0} =
[V^{h,2}]^d \times V^{h,0} \times V^{h, \mathfrak{s}}$ and $\mathfrak{s} \ge
1$. See~\cite{auricchio2017mixed,boffi2013mixed} for example.
\end{itemize}
\section{Implementation}
\label{sec:implementation}


Despite homogenisation, writing the code for the finite element system
assembly of~\cref{eq:form_boussinesq_total} is made particularly difficult and
verbose by the Nitsche boundary terms. The complexity of the implementation is
spectacularly prone to human error. To combat this issue we represent the
finite element formulation using the computational symbolic algebra package
UFL~\cite{ufl}. Combined with the core components of the FEniCS and FEniCS-X
projects~\cite{AlnaesBlechta2015a,fenicsx}, this symbolic representation is
then translated to high performance code for the assembly of the finite
element system.

The UFL offers a much more straightforward method for generating finite
element code of standard problems. However, the extra boundary terms found
in~\cref{eq:form_boussinesq_total} are still verbose, even in this friendlier
computational framework. 

Prior work has shown the use of the FEniCS project for the automatic
formulation of DG finite element formulations in the library
\ilc{dolfin\_dg}~\cite{nate2018}. Given that the DG method can be
considered a generalisation of Nitsche's method, we can exploit the
\ilc{dolfin\_dg} library to automatically formulate and generate the code
required by the Nitsche boundary terms.

We summarise the implementation details in this section, highlighting the
benefits of the mathematical homogenisation we have shown
in~\cref{sec:weak_imposition}. Python scripts will be shown to elucidate
features specific for the formulation of the exterior integral terms. However,
for a background of development and intricacies of the design philosophy we
refer to~\cite{nate2018}.

\subsection{Automatic finite element \emph{formulation--formulation} using
FEniCS and \ilc{dolfin\_dg}}

FEniCS and FEniCS-X provide a succinct way of describing and solving the
Stokes system, see the many documented examples in~\cite{fenics:book}. We
illustrate one case using a manufactured solution in \cref{code:stokes_full}.
The full details of this case will be discussed in \cref{sec:mms}. For now we
note that it follows a standard FEniCS workflow; describing the mesh and
geometry where $n = \ilc{n}$, the finite element functionspace
(here the lowest-order Taylor--Hood element) $W^{h,\text{TH}} =
[V^{h,2}]^d \times V^{h,1} = \ilc{W}$, and expressions for the known
manufactured solution \ilc{u\_soln} and \ilc{p\_soln}, and an initial
guess \ilc{U} for solution by Newton's iterative method.

To describe the Stokes system, and weakly impose slip boundary conditions
using Nitsche's method, the crucial function to define is
$\mathcal{F}^v(u, \nabla u) = \ilc{F\_v}$:
\begin{lstlisting}[language=Python]
def F_v(u, grad_u, p_local=None):
    if p_local is None:
        p_local = p
    return 2 * eta(u) * sym(grad_u) - p_local * Identity(2)
\end{lstlisting}
\ilc{doflin\_dg} requires this to have two arguments: $u = \ilc{u}$ and $\nabla u =
\ilc{grad\_u}$.  In this example it also depends on a function defining the
viscosity $\eta(u) = \ilc{eta(u)}$ and the pressure $p = \ilc{p}$.

The finite element solution variables $u = \ilc{u}$ and $p = \ilc{p}$
and respective test functions $v = \ilc{v}$ and $q = \ilc{q}$
corresponding to velocity and pressure are defined such that the variational
formulation may be written. \ilc{F\_v} is used to define the non-linear
residual for the Stokes system $\mathcal{N}_\text{Stokes} =
\mathcal{N}_\text{momentum} + \mathcal{N}_\text{mass} = \ilc{N}$.  We also
use \ilc{F\_v} to define the momentum source $f = \ilc{f}$ and the
tangential component of the Neumann boundary condition $g_{\tau,1} =
\ilc{g\_tau}$ based on the manufactured solution, where
$P_\tau(v) = \ilc{tangential\_proj(v, n)}$.

To complete the description of the problem the non-linear residual requires
the addition of the Nitsche boundary terms
\cref{eq:nitsche_form,eq:nitsche_slip_form}.  Without finite element
\emph{formulation--formulation} this would be a tedious and error--prone task.
Instead \ilc{dolfin\_dg} uses the viscous flux returned by \ilc{F\_v} to
compute the homogeneity tensor $G$ and formulate the boundary terms
automatically. For a generic problem a utility class exists for this
purpose:
\begin{lstlisting}[language=Python]
nitsche = NitscheBoundary(F_v, u, v, sigma, DGFEMClass)
\end{lstlisting}
\ilc{F\_v}, \ilc{u} and \ilc{v} are the viscous flux, unknown
solution and test functions as before. Two optional keyword arguments are also
allowed. \ilc{sigma} is the interior penalty parameter (defaulting to
\ilc{sigma}$=C_\text{IP} \frac{\ell^2}{h_F}$ where $C_\text{IP} = 20$).  
\ilc{DGFemClass} refers to one of the existing numerical flux formulation
classes, e.g. \ilc{DGFemSIPG} (the default), \ilc{DGFemNIPG} and \ilc{DGFemBO}
corresponding to SIPG, non-symmetric IPG and Baumann Oden methods,
respectively. As stated earlier in \cref{sec:weak_imposition} we are not
limited to the SIPG formulation.

The instance of the \ilc{NitscheBoundary} class can then be used to add the
Nitsche boundary terms to the non-linear residual:
\begin{lstlisting}[language=Python]
N += nitsche.nistche_bc_residual(u_soln, ds)
\end{lstlisting}
where \ilc{ds} is the integration measure defined on the exterior boudary of
the domain.

Rather than use the generic \ilc{NitscheBoundary} class, in
\cref{code:stokes_full} we use a further abstraction provided by
\ilc{dolfin\_dg} for the application of slip boundary conditions to the Stokes
system:
\begin{lstlisting}[language=Python]
stokes_nitsche = StokesNitscheBoundary(F_v, u, p, v, q)
N += stokes_nitsche.slip_nitsche_bc_residual(u_soln, g_tau, ds)
\end{lstlisting}
In addition to including the boundary terms arising in the momentum equation
\cref{eq:nitsche_slip_form}, \ilc{StokesNitscheBoundary} also includes the 
specialised treatment of the continuity formulation \cref{eq:cont_fem_form}. 
\ilc{StokesNitscheBoundary} can also take the same optional keyword arguments
as \ilc{NitscheBoundary}, \ilc{sigma} and \ilc{DGFEMClass}. Although this
implementation is designed for the incompressible Stokes formulation, this
is not a restriction, see for example~\cite{nate_phd}.

Now fully formulated the solution $(u, p) = \ilc{U}$ such that
$\mathcal{N}_\text{Stokes} = \ilc{N} = 0$ may be found, as shown in
\cref{code:stokes_full}. We invite the reader to examine many other examples
exhibited in the \ilc{dolfin\_dg} repository.

\begin{table}
\begin{lstlisting}[language=Python, caption={Example Stokes solver code using
FEniCS and \ilc{dolfin\_dg} for the manufactured solution case 1 exhibited in
\cref{sec:mms}.},captionpos=b , label={code:stokes_full}]
from dolfin import *
from dolfin_dg import StokesNitscheBoundary, tangential_proj

# Geometry
mesh = UnitSquareMesh(32, 32)
n = FacetNormal(mesh)

# Function space
We = MixedElement([VectorElement("CG", mesh.ufl_cell(), 2),
                   FiniteElement("CG", mesh.ufl_cell(), 1)])
W = FunctionSpace(mesh, We)

# Manufactured solution
u_soln = Expression(("2*x[1]*(1.0 - x[0]*x[0])",
                     "-2*x[0]*(1.0 - x[1]*x[1])"),
                    degree=4, domain=mesh)
p_soln = Constant(0.0)

# Construct an initial guess with no singularity in eta(u)
U = interpolate(Expression(("x[1]", "x[0]", "0.0"), degree=1), W)

# Viscosity model
def eta(u):
    return 1 + sqrt(inner(grad(u), grad(u)))**-1

# Viscous flux operator
def F_v(u, grad_u, p_local=None):
    if p_local is None:
        p_local = p
    return 2 * eta(u) * sym(grad_u) - p_local * Identity(2)

# Variational formulation
u, p = split(U)
v, q = split(TestFunction(W))

f = -div(F_v(u_soln, grad(u_soln), p_soln))
g_tau = tangential_proj(F_v(u_soln, grad(u_soln), p_soln) * n, n)

N = inner(F_v(u, grad(u)), grad(v)) * dx - dot(f, v) * dx \
    + div(u) * q * dx

# Slip boundary conditions
stokes_nitsche = StokesNitscheBoundary(F_v, u, p, v, q)
N += stokes_nitsche.slip_nitsche_bc_residual(u_soln, g_tau, ds)

solve(N == 0, U)
\end{lstlisting}
\end{table}

\subsection{Computational tools}

With the computational symbolic representation of the finite element problem
in place we use DOLFIN to facilitate the computation of the finite element
solution~\cite{logg:2010}. Additionally we make use of DOLFIN-X of the
FEniCS-X sequel project, primarily for its features of block matrix assembly
and high order (curved) facet representation~\cite{fenicsx}.

The underlying linear algebra systems are solved using the data structures and
algorithms provided in the portable extensible toolkit for scientific
computation (PETSc)~\cite{petsc-web-page,petsc-user-ref,petsc_efficient}.
Specifically, the direct solver MUMPS~\cite{mumps} is used for all 2D
computations. For 3D computations we employ HYPRE-BoomerAMG~\cite{henson:2002}
and PETSc's geometric algebraic multigrid (GAMG) preconditioner.

\section{Numerical experiments}
\label{sec:experiments}

In this section numerical experiments are devised for the finite element
formulation of the Stokes subsystem of the Boussinesq approximation with free
slip boundary conditions. A manufactured solution is first considered as
presented in Urquiza~et~al.~\cite{urquiza2014}. Secondly, established
benchmark problems in the literature are chosen. We consider the isoviscous
and temperature dependent viscosity steady state convection benchmarks
constructed by Blankenbach~et~al.~\cite{blankenbach1989}, in addition to the
steady state convection benchmarks devised by Tosi~et~al.~\cite{tosi2015}. The
benchmarks devised by Tosi~et~al. are particularly interesting given the
highly nonlinear temperature and strain rate dependent viscosity model. These
benchmarks are composed from rectangular geometries whose boundaries are
aligned with the Cartesian coordinate system. Clearly it is not necessary to
use a weakly imposed free slip condition in this setting; however they provide
familiarity and direct comparison of solution functionals. Finally we consider
a subduction zone model as exhibited in~\cite{vankeken2008}. Here we examine
the use of the free slip boundary condition on a downward subducting slab.
Additionally we investigate the viability of Krylov subspace iterative methods
for the solution of the finite element system.

\subsection{Example 1: Manufactured solution}
\label{sec:mms}

This manufactured solution example is a reproduction of the numerical
experiments examined in~\cite{urquiza2014}, however, with a modification to the
viscosity generating a nonlinear formulation (see~\cref{eq:mms_visc}).

Let the domain $\Omega \coloneqq (-1, 1)^2$. The
domain is partitioned into a sequence of nested conforming meshes containing
$2\times N^2$, $N \in \mathbb{N}$, triangles which are formed by bisecting
squares of side length $N^{-1}$.

\begin{figure}
\centering
\begin{subfigure}{.5\textwidth}
  \centering
	\begin{tikzpicture}
	\begin{axis}[
	axis equal image, 
	domain=-0.95:0.95,
	xmin=-1.3, xmax=1.3,
	ymin=-1.3, ymax=1.3,
	view={0}{90},
	xtick=\empty,
	ytick=\empty,
	xlabel={$x$},
	ylabel={$y$},
	axis lines=middle]
	\addplot3[blue, quiver={u={2*y*(1-x^2)}, v={-2*x*(1-y^2)}, scale arrows=0.15}, -stealth,samples=15] {0};
	\addplot[samples=100, line width=2pt, domain=0:2*pi] 
	({cos(deg(x))}, {sin(deg(x))});
	\addplot[black, line width=2pt] coordinates {(-1, -1) (-1, 1) (1, 1) (1, -1) (-1, -1)};
	\end{axis}
	\end{tikzpicture} \\
  Case 1
\end{subfigure}%
\begin{subfigure}{.5\textwidth}
  \centering
	\begin{tikzpicture}
	\begin{axis}[
	axis equal image, 
	domain=-0.95:0.95,
	xmin=-1.3, xmax=1.3,
	ymin=-1.3, ymax=1.3,
	view={0}{90},
	xtick=\empty,
	ytick=\empty,
	xlabel={$x$},
	ylabel={$y$},
	axis lines=middle]
	\addplot3[blue, quiver={u={-y*sqrt(x^2 + y^2)}, v={x*sqrt(x^2 + y^2)}, scale arrows=0.15}, -stealth,samples=15] {0};
	\addplot[samples=100, line width=2pt, domain=0:2*pi] 
	({cos(deg(x))}, {sin(deg(x))});
	\addplot[black, line width=2pt] coordinates {(-1, -1) (-1, 1) (1, 1) (1, -1) (-1, -1)};
	\end{axis}
	\end{tikzpicture} \\
  Case 2
\end{subfigure}
\caption{Analytical solutions of the manufactured solution experiments in
numerical examples 1 and 2. Overlaid are the square $\Omega = (-1, 1)^2$
and elliptical $\Omega = \{(x, y) : x^2 + \frac{y^2}{(1 +
\varepsilon_y)^2} < 1\}$ domains, where $\varepsilon_y = 0$, used in examples
1 and 2, respectively.}
\label{fig:mms_anal_soln}
\end{figure}
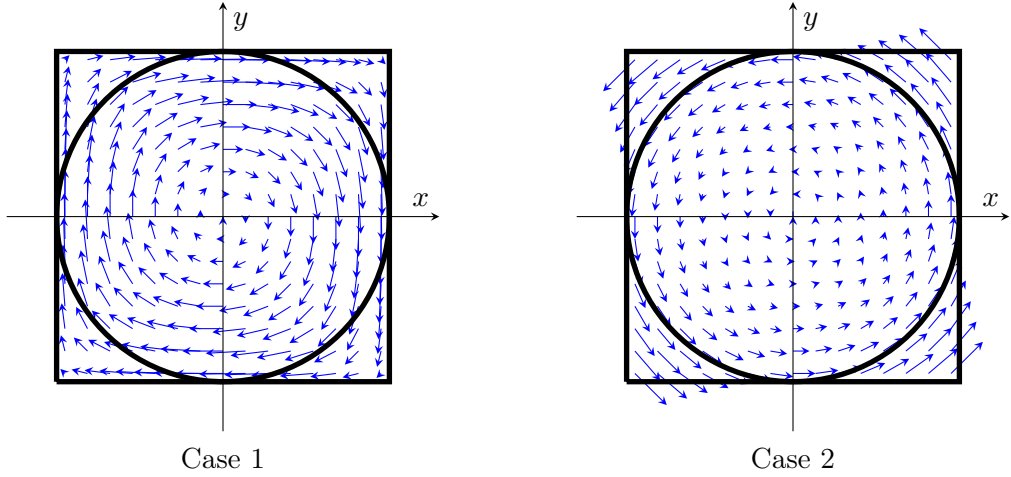

The finite element solution of the Stokes
system~\cref{eq:stokes_vel,eq:stokes_cont} is computed where the
\emph{a priori} known solution is prescribed by the following two cases:
\begin{align}
\text{Case 1: } u &= \left(2 y (1 - x^2), -2 x(1 - y^2)\right)^\top \nonumber, \\
\text{Case 2: } u &= \left(-y \sqrt{x^2 + y^2}, x \sqrt{x^2 + y^2}\right)^\top. \nonumber
\end{align}
In both cases the pressure is given by $p=0$. The viscosity is chosen such
that
\begin{equation}
\eta = \left(1 + \sqrt{\varepsilon(u) :\varepsilon(u)}\right)^{-1}. \label{eq:mms_visc}
\end{equation}{}
The boundary conditions are selected such that $\Gamma_S = \partial\Omega$.
$u_S$, $g_{\tau,1}$ and $f$ are computed accordingly from the known solution.
The analytical solutions in each case are shown in \cref{fig:mms_anal_soln}.

\begin{figure}
\centering
\begin{subfigure}{.5\textwidth}
  \centering
  \includegraphics[width=1.\linewidth]{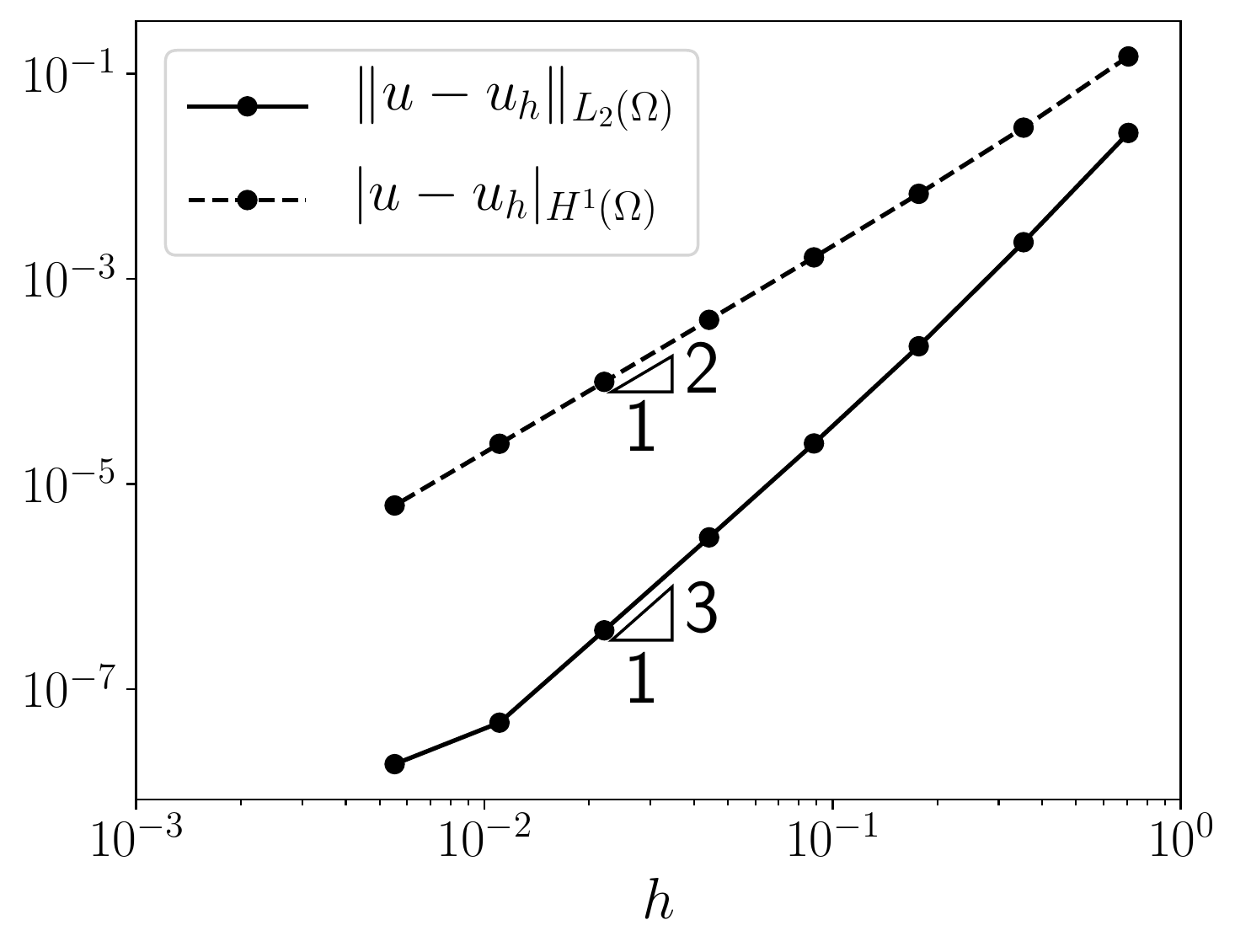} \\
  (a)
\end{subfigure}%
\begin{subfigure}{.5\textwidth}
  \centering
  \includegraphics[width=1.\linewidth]{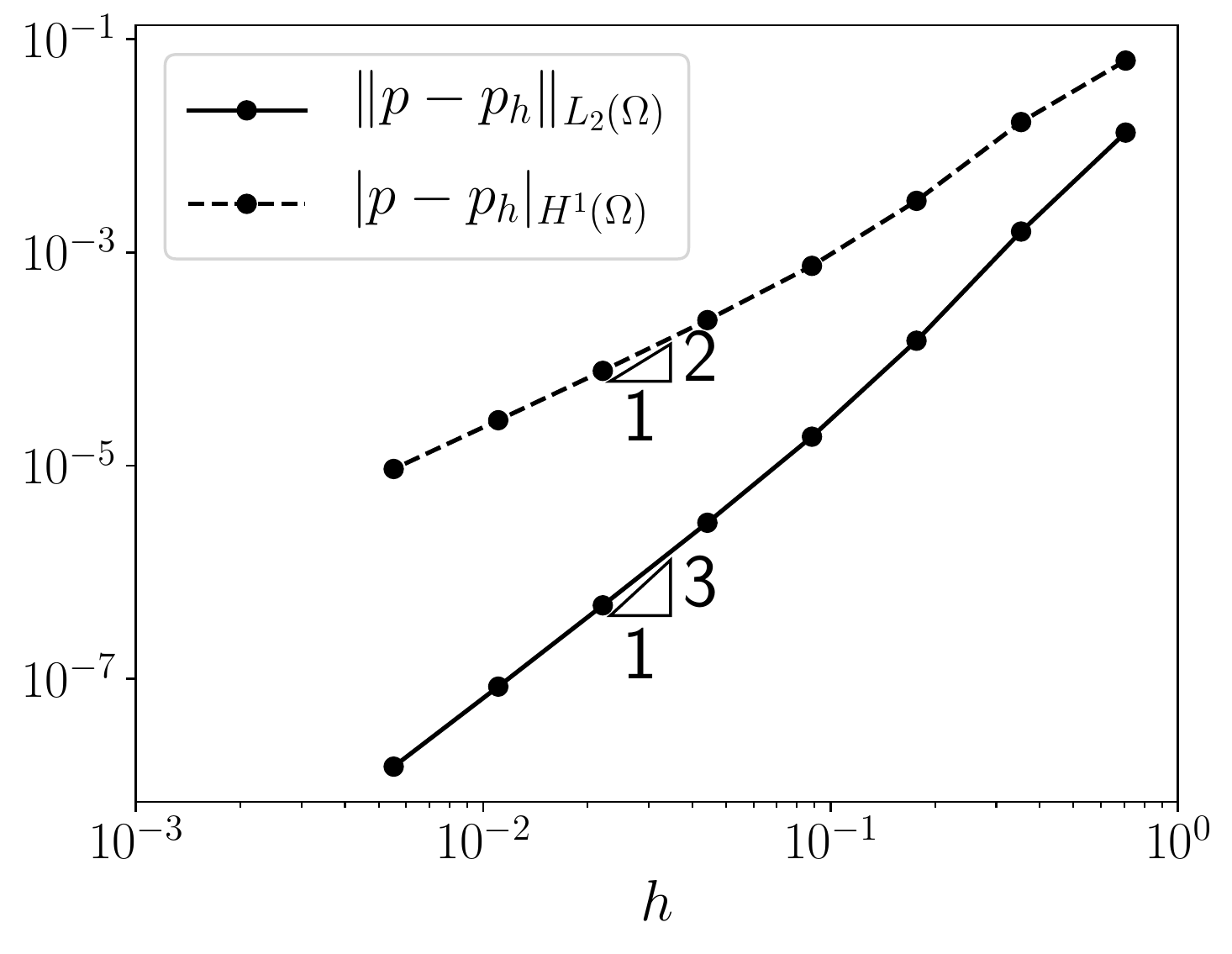} \\
  (b)
\end{subfigure}
\begin{subfigure}{.5\textwidth}
  \centering
  \includegraphics[width=1.\linewidth]{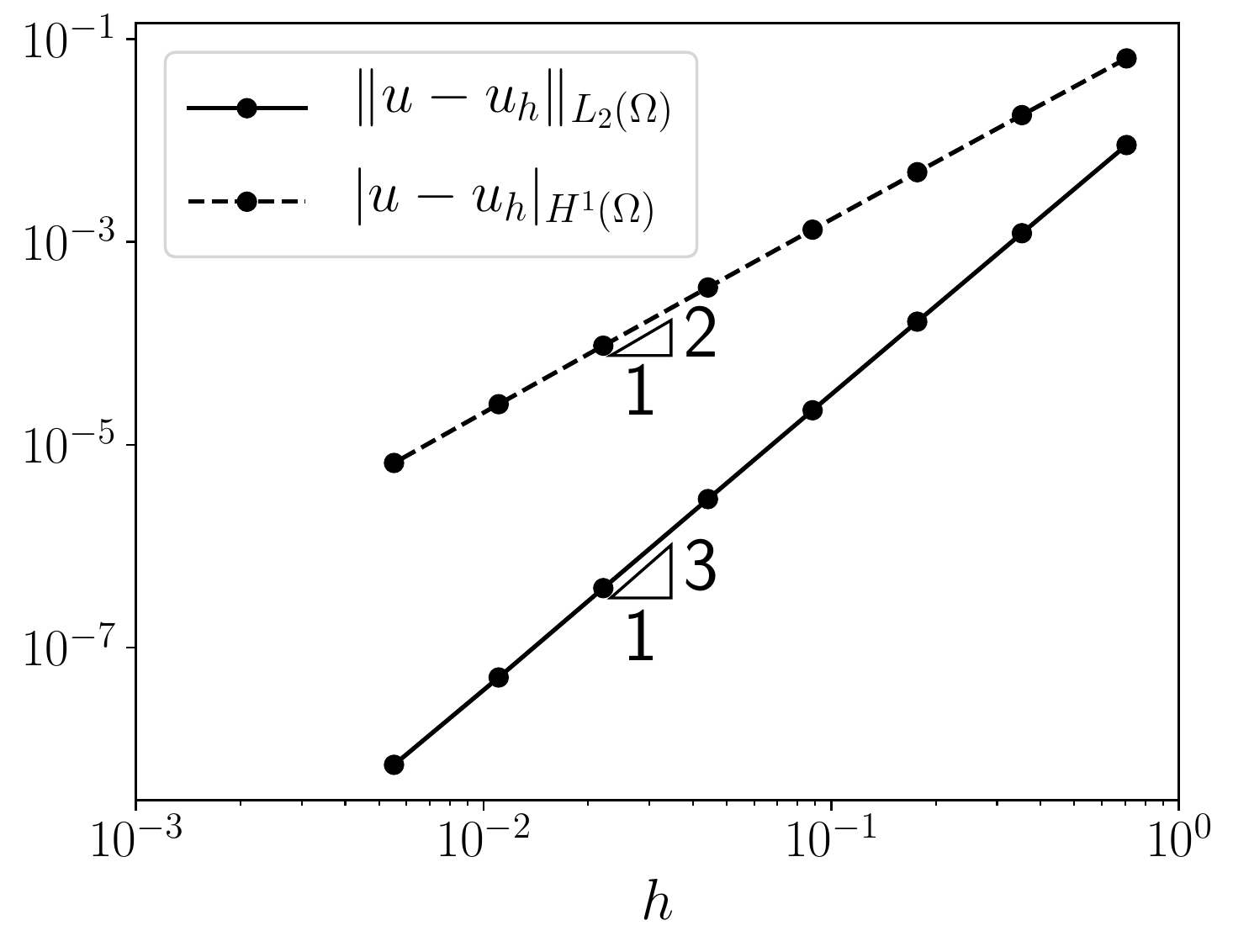} \\
  (c)
\end{subfigure}%
\begin{subfigure}{.5\textwidth}
  \centering
  \includegraphics[width=1.\linewidth]{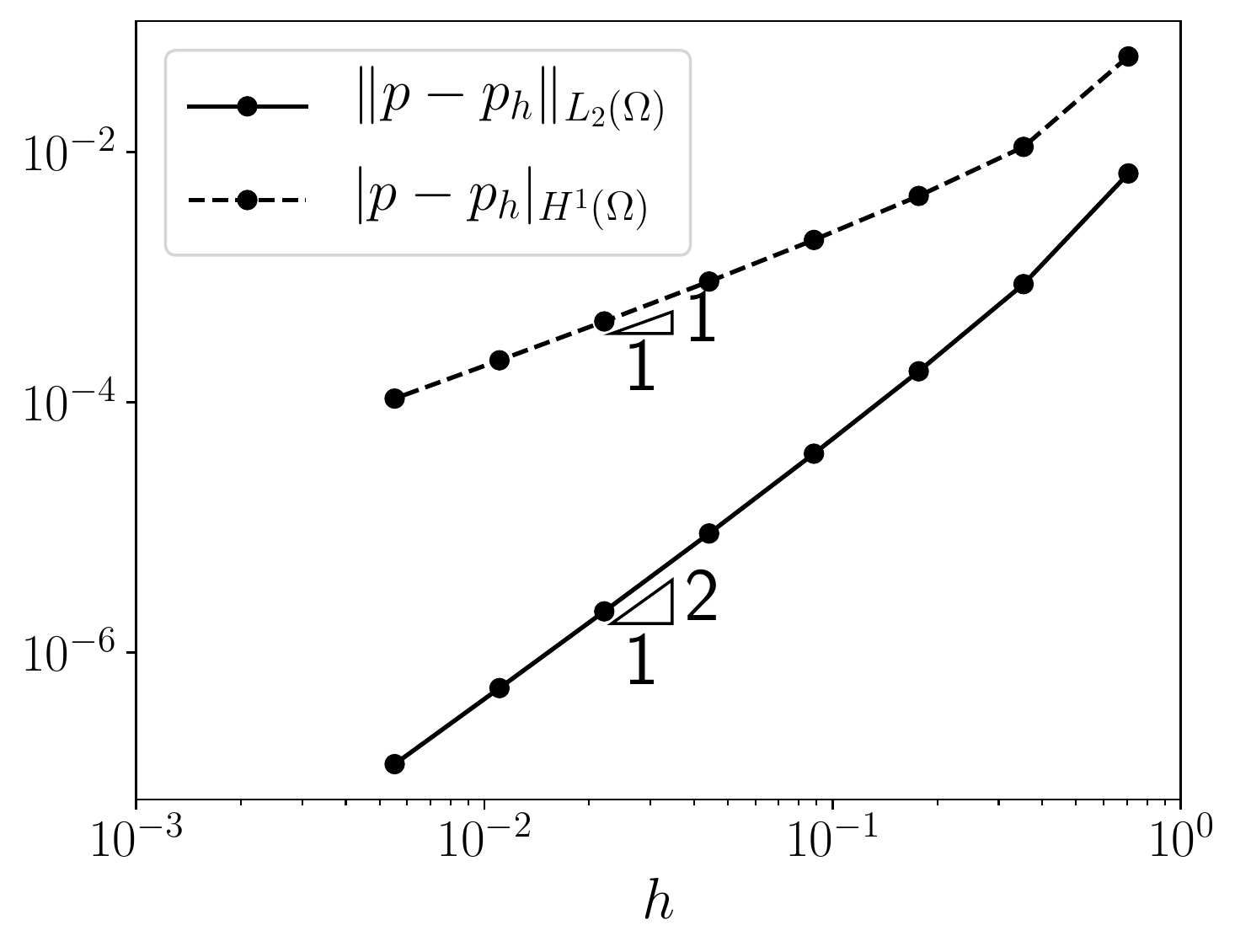} \\
  (d)
\end{subfigure}
\caption{Example 1: Convergence of the FEM Stokes system with $h$-refinement. Case~1: (a)
velocity solution, (b) pressure solution. Case~2: (c) velocity solution, (d) pressure solution.}
\label{fig:mms_conv_rates}
\end{figure}

Finite element error convergence rate results are presented
in~\cref{fig:mms_conv_rates}. Note that these results compare well with those
presented in~\cite{urquiza2014}. In both cases we recover optimal convergence
rates with the Taylor-Hood discretisation.

\subsection{Example 2: Manufactured solution - Babu\v{s}ka's paradox}

Babu\v{s}ka's paradox is a well known problem in computational solid
mechanics~\cite{babuvska1990}. It states that with simple support boundary
conditions the solution of the Kirchoff-Love plate equation (biharmonic
equation) is not the same as the solution of same equations posed on a
polygonal domain in the limit approaching the unit disc~\cite{babuvska1990}.
This is particularly pertinent in our experiments given that the solution of
the Stokes system with free slip boundary conditions may be reformulated in
terms of the stream function solution of the biharmonic equation. For more
details on Babu\v{s}ka's paradox in the context of the Stokes system and a
review of methods available to alleviate its consequences we refer
to~\cite{urquiza2014}.

Babu\v{s}ka's paradox is problematic given that the finite element
approximation of the Stokes solution is at the mercy of the variational crime
$\Omega \neq \Omega_h$. By this we mean that a piecewise polynomial
representation of the exterior facets of the mesh may only be an approximation
of the disk. In this section we will examine \emph{a priori} error convergence
properties as the computational domain's fidelity approaches the unit disk. We
consider the case where the polynomial representation of the exterior facets
is piecewise quadratic.

Here we repeat the manufactured solution numerical experiments of the previous
section on the ellipse $\Omega \vcentcolon= \{(x, y) : x^2 + \frac{y^2}{(1 +
\varepsilon_y)^2} < 1\}$. Note that the domain is the unit disc in the case
that $\varepsilon_y = 0$. By choosing parameters $\varepsilon_y \ll 1$, we can
investigate the impact of Babu\v{s}ka's paradox on the solvability of the
problem. The analytical solutions in each case are shown in
\cref{fig:mms_anal_soln}.

\begin{figure}
\centering
\begin{subfigure}{.5\textwidth}
  \centering
  \includegraphics[width=1.\linewidth]{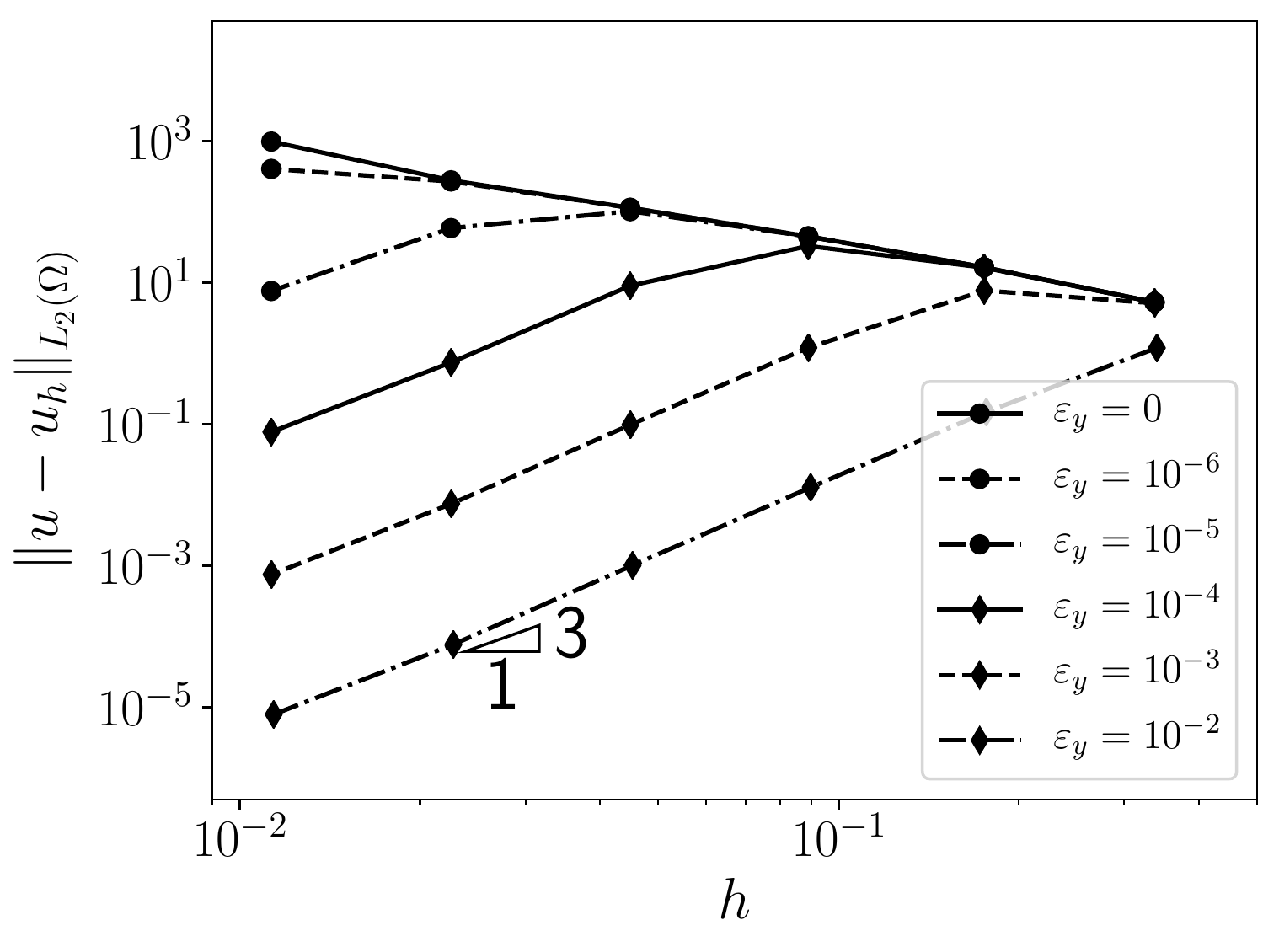} \\
  (a)
\end{subfigure}%
\begin{subfigure}{.5\textwidth}
  \centering
  \includegraphics[width=1.\linewidth]{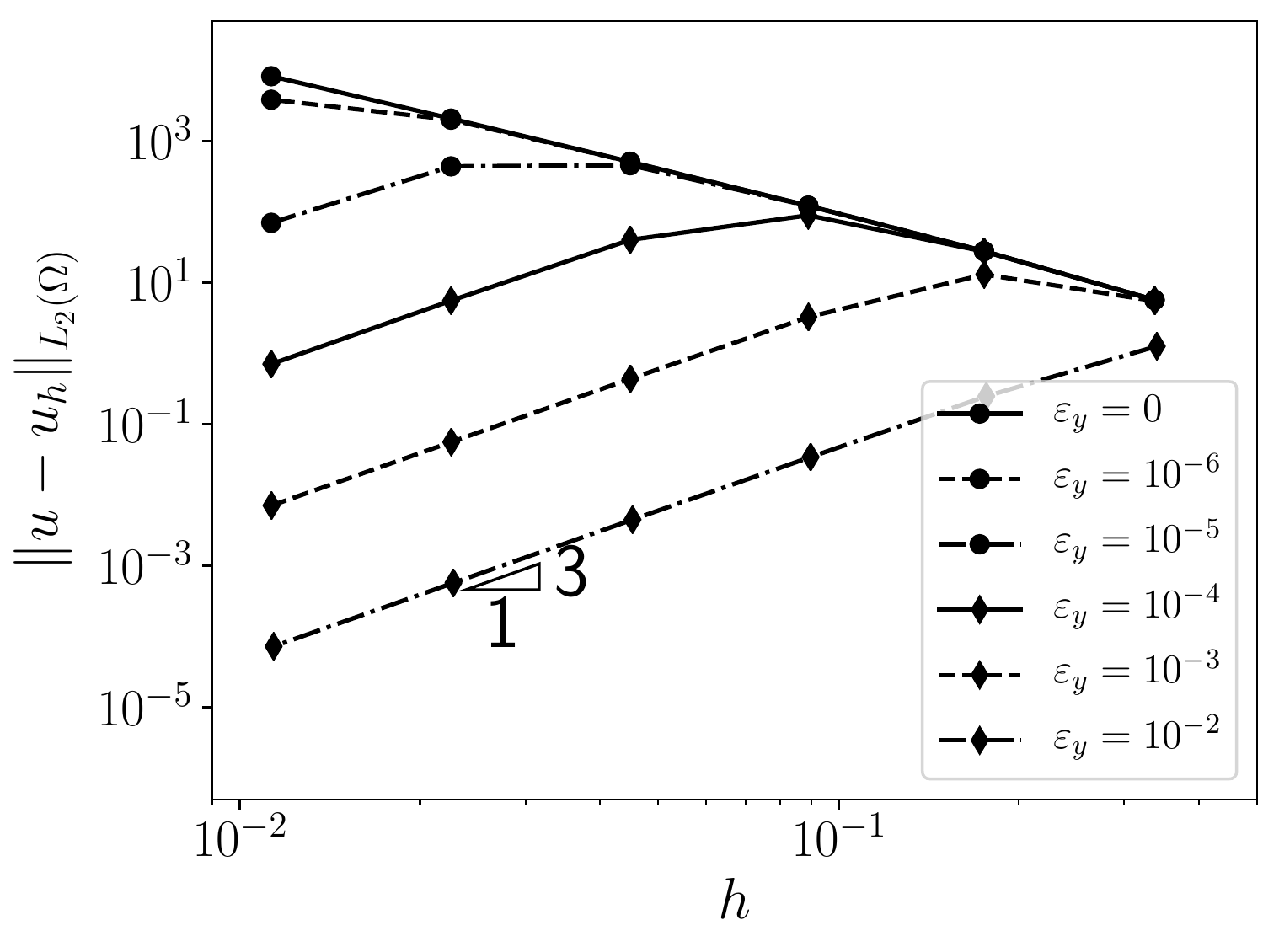} \\
  (b)
\end{subfigure}
\caption{Example 2: Convergence of the FEM Stokes system in the ellipse with
$h$-refinement and free slip boundary conditions. Note that as $\varepsilon_y
\rightarrow 0$, the finite element velocity solution does not converge to the
true velocity solution. (a) Case 1, (b) Case 2.}
\label{fig:babuska_conv_rates}
\end{figure}

The results are shown in \cref{fig:babuska_conv_rates}. In the case that
$\varepsilon_y = 0$, clearly the system does not converge to the true
solution, a demonstration of the perplexity of Babu\v{s}ka's paradox. In the cases
that $\varepsilon_y > 0$ one observes that as the approximation of the ellipse
is more precise (i.e. we diverge from the unit disc), the finite element
approximation error indicates convergence to the true solution at an optimal
rate.

\subsection{Example 3: Steady state isoviscous and nonlinear convection}
\label{sec:blankenbach_results}

\begin{table}
\centering
\begin{subtable}{0.5\textwidth}
\centering
\scalebox{0.85}{
\begin{tabular}{lrrrrr}
Case & $\mathrm{Ra}$ & $\Delta \eta_T$ & $\Delta \eta_z$ & $L$  & $\eta$ \\ \hline
1    & \num{e4}      & \num{0}         & \num{0}         & 1    & $\eta_\text{lin}$     \\
2    & \num{e5}      & \num{0}         & \num{0}         & 1    & $\eta_\text{lin}$     \\
3    & \num{e6}      & \num{0}         & \num{0}         & 1    & $\eta_\text{lin}$     \\
4    & \num{e4}      & \num{e3}        & \num{0}         & 1    & $\eta_\text{lin}$     \\
5    & \num{e4}      & \num{16384}     & \num{64}        & 2.5  & $\eta_\text{lin}$
\end{tabular}
}
\caption{Blankenbach~\cite{blankenbach1989} benchmark cases.}
\end{subtable}%
\begin{subtable}{0.5\textwidth}
\centering
\scalebox{0.85}{
\begin{tabular}{lrrrrrrr}
Case & $\mathrm{Ra}$ & $\Delta \eta_T$ & $\Delta \eta_z$ &  $\eta^*$ & $\sigma_Y$ & $L$ & $\eta$ \\ \hline
1    & \num{e2}      & \num{e5}         & \num{1}        &           &            & 1   & $\eta_\text{lin}$ \\
2    & \num{e2}      & \num{e5}         & \num{1}        & \num{e-3} & \num{1}    & 1   & $\eta_\text{total}$ \\
3    & \num{e2}      & \num{e5}         & \num{10}       &           &            & 1   & $\eta_\text{lin}$ \\
4    & \num{e2}      & \num{e5}         & \num{10}       & \num{e-3} & \num{1}    & 1   & $\eta_\text{total}$
\end{tabular}
}
\caption{Tosi~\cite{tosi2015} benchmark cases.}
\end{subtable}
\caption{Benchmark cases and their corresponding viscosity model parameters.}
\label{tab:blankenbach_tosi_params}
\end{table}

The first numerical example of a geophysical Boussinesq system is the steady
state benchmark problems collated by Blankenbach~et~al.~\cite{blankenbach1989}
and additional similar benchmarks collated by Tosi~et~al.~\cite{tosi2015}.
Here, mantle convection is modelled in a rectangular cell of length $L$ and
height $H=1$ such that $\Omega := (0,
L) \times (0, H)$. The domain is meshed into $2 \times \mathfrak{m} \times \mathfrak{n}$ regular
triangle elements where $\mathfrak{m}$ and $\mathfrak{n}$ are the number of bisected squares in
the $x$ and $y$ directions, respectively. On these meshes we compute the
finite element approximation of the Boussinesq
system~\cref{eq:stokes_vel,eq:stokes_cont,eq:heat}. The source term in the
Stokes system is given by
\begin{equation}
f = \mathrm{Ra} \: T \hat{k},
\end{equation}
where $\mathrm{Ra}$ is the dimensionless Rayleigh number and $\hat{k} = (0,
1)^\top$ is the unit vector pointing in the direction of buoyancy.
The viscosity $\eta$ given in terms of
\begin{align}
\eta_\text{lin} &= e^{- \ln(\Delta \eta_T) T + \ln(\Delta \eta_z) z}, &&
\eta_\text{plast} &= \eta^* + \frac{\sigma_Y}{\sqrt{\varepsilon(u) : \varepsilon(u)}}, &&
\eta_\text{total} &= 2 \del{ \eta_\text{lin}^{-1} + \eta_\text{plast}^{-1}}^{-1}, \label{eq:tosi_visocisty}
\end{align}
where $z = 1-y$ is the depth from the lid (located at $z=0$) and
$\Delta\eta_T$ and $\Delta\eta_z$ are constant parameters. $\sigma_Y$ is the
yield stress, $\eta^* \ll 1$ is a constant and $\sqrt{\varepsilon(u) :
\varepsilon(u)}$ is the second invariant of the rate of strain tensor. The
non--dimensional thermal diffusivity $\kappa = 1$ and heat source $Q = 0$ in
all cases.

The boundary conditions require that a free slip boundary condition is
applied on $\Gamma_S = \partial\Omega$ 
and that the temperature at the base and top of the cell is
prescribed to be $T|_{y=0} = 1$ and $T|_{y=H} = 0$. Finally $\nabla T
\cdot n = 0$ on the left $x=0$ and right $x=L$ boundaries.

Five steady state benchmark cases are examined from the work by
Blankenbach~et~al.~\cite{blankenbach1989} and four steady state benchmark
cases showcased by Tosi~et~al.~\cite{tosi2015}. The parameters employed in
these benchmarks are shown in~\cref{tab:blankenbach_tosi_params},
respectively. The key difference between the two benchmarks is the viscosity
model. The Blankenbach~et~al.~benchmarks focus on linear, temperature and
depth dependent viscosity, whereas the Tosi~et~al.~benchmark includes a
visco-plastic model dependent on temperature and rate of strain, giving rise
to a nonlinear system of equations.

To find the finite element solution we initially use a fixed point iteration
between the Stokes and temperature subsystems. After around \num{10} fixed
point iterations, the approximation is supplied as an initial guess to a
Newton solver for subsequent solution. Convergence is satisfied if the
absolute value of the $2-$norm of the finite element residual vector is less
than~\num{e-10}. See~\cite{Vynnytska2013} for an example of the fixed point
method in this context.

\begin{figure}
\centering
\begin{subfigure}{.5\textwidth}
  \centering
  \includegraphics[width=1.\linewidth]{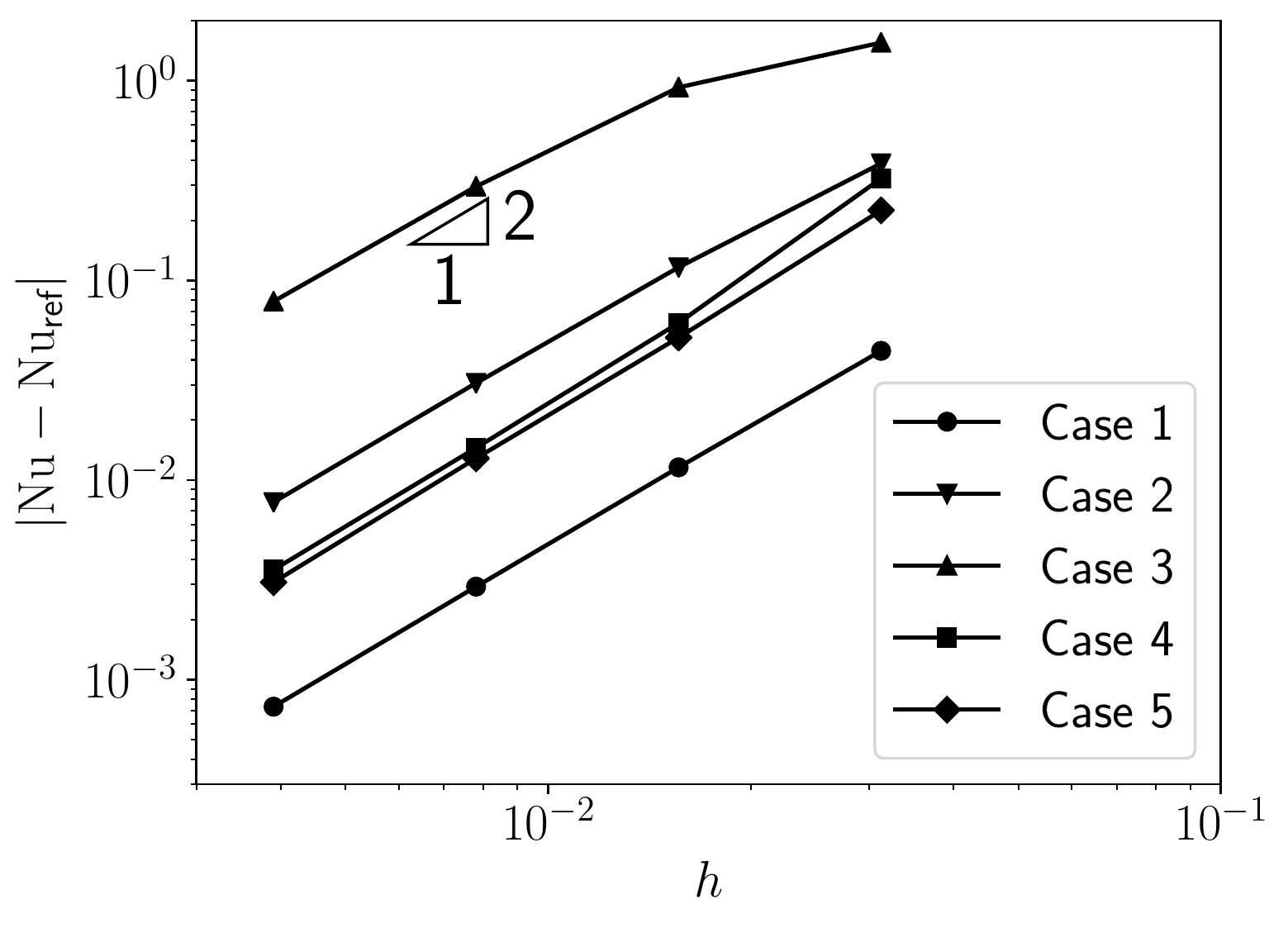} \\
  (a)
\end{subfigure}
\caption{Example 3: Convergence of the computed top Nusselt number approximation
in the Blankenbach~et~al.~\cite{blankenbach1989} benchmark using Nitsche's
method to enforce the free slip boundary condition. See
\cref{tab:blankenbach_tosi_params} for a list of parameters. In case~5 of the
Blankenbach~et~al. benchmarks, $\mathfrak{m} = \frac{5}{2} \mathfrak{n}$. The
reference values $\mathrm{Nu}_\text{ref}$ used are obtained from
\cite{Vynnytska2013}.}
\label{fig:blankenbach_convergence}
\end{figure}

A complete list of computed functionals and their tabulation with mesh
refinement are given in \cref{sec:blankenbach_tosi_tables} where the free slip
boundary conditions are enforced in both the strong and weak sense for
comparison. For brevity, here we show the computed top Nusselt number
\begin{equation}
\mathrm{Nu}_\text{top} = 
\frac{\int_{\Gamma_\text{top}} \nabla T \cdot n \dif{s}}
{\int_{\Gamma_\text{bottom}} T \dif{s}}
\end{equation}
from the benchmark cases solved with weakly imposed free slip boundary
conditions in \cref{fig:blankenbach_convergence} and \cref{tab:tosi_results},
respectively. Here $\Gamma_\text{top}$ and $\Gamma_\text{bottom}$ correspond
to the exterior boundaries at $y = H$ and $y = 0$, respectively. All
functionals converge in agreement with the literature. Furthermore, the
results computed from the strong and weak imposition of the boundary
conditions compare favourably as shown in \cref{sec:blankenbach_tosi_tables}.

\begin{table}
\centering
\begin{subtable}{\textwidth}
  \centering
  \small
	\begin{tabular}{lrllll}
	Code & $\mathfrak{m}$ & Case 1   & Case 2   & Case 3   & Case 4   \\ \hline
	This work & 32  & \num{3.420410} & \num{8.696873} & \num{3.034962} & \num{6.711525} \\
	& 64  & \num{3.424483} & \num{8.603156} & \num{3.035024} & \num{6.642305} \\
	& 128 & \num{3.424629} & \num{8.571444} & \num{3.034917} & \num{6.622504} \\
	& 256 & \num{3.424609} & \num{8.562694} & \num{3.034883} & \num{6.617284} \\ \hline
	StagYY & 128 & \num{3.419} & \num{8.5491} & \num{3.03025} & \num{6.61082} \\
	Fluidity & 128 & \num{3.4253} & \num{8.5693} & \num{3.0399} & \num{6.6401} \\
	ASPECT & 64 & \num{3.4305} & \num{8.5758} & \num{3.0371} & \num{6.6249}
	\end{tabular}
\end{subtable}
\caption{Example 3: Computed top Nusselt number values $\mathrm{Nu}_\text{top}$
using Nitsche's method to enforce the free slip boundary conditions of the
benchmark~\cite{tosi2015}. Reference values from finite element codes are
selected from~\cite{tosi2015} for comparison. See
\cref{tab:blankenbach_tosi_params} for a list of parameters.}
\label{tab:tosi_results}
\end{table}

\subsection{Example 4: Subduction zone}
\label{sec:subduction_zone}

The subduction zone model in this section comprises a downward sloping slab
incident to an overriding plate lying below the lithosphere. We will reproduce
the numerical benchmark undertaken in~\cite{vankeken2008}. Note however, that
in this benchmark the velocity at the interface between the downward sloping
slab and the overriding plate is not prescribed in the sense of zero
penetration, but rather a fixed velocity boundary. In this section we take a
departure from this requirement of the benchmark in favour of a comparison of
prescribed velocity data and a free slip (no penetration) boundary condition
between the subducting slab and overriding plate using Nitsche's method. We
validate our experiments by reproducing the numerical experiment cases 1c, 2a
and 2b undertaken in the benchmark~\cite{vankeken2008}.

\subsubsection{Geometry}

\begin{figure}
\centering
\begin{subfigure}{.75\textwidth}
  \resizebox{\linewidth}{!}{
  \begin{tikzpicture}[scale=0.5]
	\begin{pgfonlayer}{nodelayer}
		\node [style=none] (1) at (0, -12) {};
		\node [style=none] (2) at (13, -12) {};
		\node [style=none] (3) at (13, 0) {};
		\node [style=none] (4) at (12, -12) {};
		\node [style=none] (5) at (0, 0) {};
		\node [style=none] (6) at (2, -7.25) {$u = u_{\text{inlet}}$};
		\node [style=none] (7) at (8.5, -5.25) {$\Gamma_\mathrm{slab}$};
		\node [style=none] (8) at (16.5, -6) {$(2 \eta \varepsilon(u) - pI) \cdot n = 0$};
		\node [style=none] (9) at (7.5, -1) {$u = 0$};
		\node [style=none] (10) at (5.25, -11.5) {$(2 \eta \varepsilon(u) - pI) \cdot n = 0$};
		\node [style=none] (11) at (-0.5, 0) {};
		\node [style=none] (12) at (-0.5, -12) {};
		\node [style=none] (13) at (-1, -6) {$Z$};
		\node [style=none] (14) at (1, -1) {};
		\node [style=none] (15) at (1.5, 0) {};
		\node [style=none] (16) at (2.25, -0.75) {$\alpha$};
		\node [style=none] (17) at (12, -12.5) {};
		\node [style=none] (18) at (13, -12.5) {};
		\node [style=none] (20) at (12.5, -13.25) {$\delta_x$};
		\node [style=none] (21) at (0, -12.75) {};
		\node [style=none] (22) at (6, -13.25) {};
		\node [style=none] (23) at (6, -13.25) {$Z (\tan \alpha)^{-1}$};
		\node [style=none] (25) at (12, -12.75) {};
		\node [style=none] (26) at (0, 0.5) {$(0, 0)$};
		\node [style=none] (27) at (15, -12) {};
		\node [style=none] (28) at (16, -12) {};
		\node [style=none] (29) at (15, -11) {};
		\node [style=none] (30) at (16.5, -12) {$x$};
		\node [style=none] (31) at (15, -10.5) {$y$};
		\node [style=none] (32) at (2, -2) {};
		\node [style=none] (33) at (13, -2) {};
		\node [style=none] (34) at (1.5, -3) {$u = \tau_r$};
		\node [style=none] (35) at (13.5, 0) {};
		\node [style=none] (36) at (13.5, -2) {};
		\node [style=none] (38) at (1.25, -1.5) {};
		\node [style=none] (39) at (1, -2.5) {};
		\node [style=none] (40) at (14.5, -1) {$Z_\mathrm{plate}$};
		\node [style=none] (41) at (8.5, -5.75) {};
		\node [style=none] (42) at (7.25, -7) {};
	\end{pgfonlayer}
	\begin{pgfonlayer}{edgelayer}
		\draw [style=thick] (5.center) to (1.center);
		\draw [style=thick] (5.center) to (3.center);
		\draw [style=thick] (3.center) to (2.center);
		\draw [style=thick] (2.center) to (4.center);
		\draw [style=thick] (4.center) to (1.center);
		\draw [style=double arrow] (11.center) to (12.center);
		\draw [in=0, out=-75, looseness=0.75] (15.center) to (14.center);
		\draw [style=double arrow] (17.center) to (18.center);
		\draw [style=double arrow] (21.center) to (25.center);
		\draw [style=arrow] (27.center) to (29.center);
		\draw [style=arrow] (27.center) to (28.center);
		\draw [style=thick] (5.center) to (14.center);
		\draw [style=thick] (14.center) to (32.center);
		\draw [style=thick] (32.center) to (4.center);
		\draw [style=thick] (32.center) to (33.center);
		\draw [style=double arrow] (36.center) to (35.center);
		\draw [style=arrow, bend left] (39.center) to (38.center);
		\draw [style=arrow, bend left, looseness=0.75] (41.center) to (42.center);
	\end{pgfonlayer}
\end{tikzpicture}
}
\end{subfigure}
\caption{The $(0, Z (\tan \alpha)^{-1} + \delta_x) \times (0, -Z)$ rectangle
subduction zone problem geometry. The subduction slab is defined by a line
sourced at $(0, 0)$ and terminating at $(Z (\tan \alpha)^{-1}, -Z)$.}
\label{fig:subduction_geometry}
\end{figure}
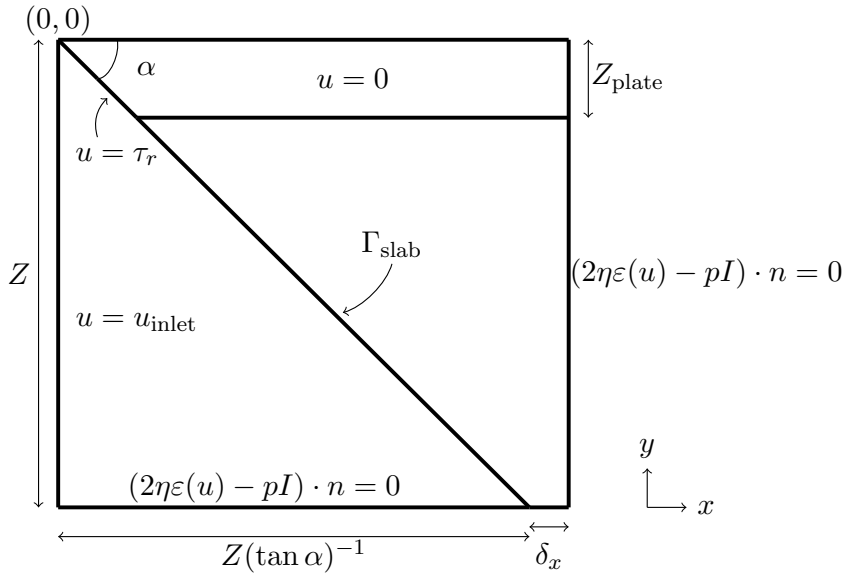

The geometry and boundary conditions of this example problem are exhibited in
\cref{fig:subduction_geometry}. Specifically, the geometry is composed of a
$(0, Z (\tan \alpha)^{-1} + \delta_x) \times (0, -Z)$ rectangle, where $Z =
\SI{600}{\kilo\metre}$ is the slab depth, $\alpha = \pi / 4$ is the dipping
angle of the slab (measured from the top of the overriding lithosphere) and
$\delta_x = \SI{60}{\kilo\metre}$ is a small extension in the $x$ direction.
The rectangle is divided by a line whose source lies at $(0, 0)$ and
terminates at $(Z (\tan \alpha)^{-1}, -Z)$. The overriding plate extends from
this subduction interface at a constant depth of $Z_\text{plate} =
\SI{50}{\kilo\metre}$ and terminating at $(Z (\tan \alpha)^{-1} + \delta_x,
-Z_\text{plate})$.

\subsubsection{Boundary conditions}

The velocity boundary conditions are as follows: The incoming slab is incident
with velocity $u = u_\text{inlet}$ on the left side boundary $x = 0$, no slip
$u = 0$ on the overriding plate $y = -Z_\text{plate}$ and a prescribed
velocity of the subducting slab in the right oriented tangential direction $u
= \tau_r$ above the the plate depth $y > -Z_\text{plate}$ where
\begin{equation}
\tau_r =
\begin{cases}
\tau & \text{if } \tau_x > 0, \\
-\tau & \text{otherwise}.
\end{cases}
\end{equation}
Natural boundary conditions are enforced on the remaining exterior boundary.
On the remaining slab interface lying below the overriding plate,
$\Gamma_\text{slab}$, we choose to enforce either a free slip boundary
condition $u \cdot n = 0$ or prescribed tangential flow $u = \tau_r$.
%

The temperature boundary conditions are as follows: on the top exterior
boundary $T(x, 0) = T_\text{s}$, on the left boundary $T(0, y) =
T_\text{left}$ and on the right exterior boundary $T(Z (\tan \alpha)^{-1} +
\delta_x, y) = T_\text{right}$, where the quantities
\begin{align}
T_\text{s} &= \num{273}, \nonumber \\
T_\text{left} &= T_\text{s} + \del{T_0 - T_\text{s}} \mathrm{erf} \del{ -y \frac{\num{e3}}{2 \sqrt{\kappa t_{\num{50}}}}}, \nonumber \\
T_\text{right} &=
\begin{cases}
T_\text{s} - \num{0.026e3} y & \text{if } y > -Z_\text{plate}, \\
T_0 & \text{otherwise}.
\end{cases}
\end{align}
Here $T_0 = \SI{1573}{\kelvin}$ is the inlet temperature, $t_{\num{50}} =
\num{50e6} \times \num{365} \times \num{24} \times \num{60}^2$ is the age of the
$\SI{50}{\mega\year}$ old plate in seconds and the thermal diffusivity $\kappa
= \SI{0.7272e-6}{\metre\squared\per\second}$.

\subsubsection{Viscosity and slab surface models}

We consider the following cases which are named corresponding with the
work~\cite{vankeken2008}:
\begin{enumerate}
\item[1c:] The linear isoviscous case $\eta = 1$.

\item[2a:] A nonlinear diffusion creep model with viscosity
\begin{equation}
\eta_\text{diff} = A_\text{diff} \exp \del{\frac{E_\text{diff}}{R T}},
\end{equation}
where $A_\text{diff} = \num{1.32043e9}$, $E_\text{diff} = \num{335e3}$ and
$R = \num{8.3145}$ from which the total viscosity
\begin{equation}
\eta = \left(\eta_\text{max}^{-1} + \eta_\text{diff}^{-1} \right)^{-1},
\end{equation}
where $\eta_\text{max} = \num{e26}$.

\item[2b:] A nonlinear dislocation creep model with viscosity
\begin{equation}
\eta_\text{disl} = A_\text{disl} \exp \left( \frac{E_\text{disl}}{n_\text{exp} R T} \right) \varepsilon_\mathrm{II}^\frac{1 - n_\text{exp}}{n_\text{exp}},
\end{equation}
where $A_\text{disl} = \num{29868.6}$, $E_\text{disl} = \num{540e3}$,
$n_\text{exp} = \num{3.5}$ and $\varepsilon_\mathrm{II} = \sqrt{\frac{1}{2}
\varepsilon (u) : \varepsilon (u)}$, from which the total
viscosity is given by
\begin{equation}
\eta = \left(\eta_\text{max}^{-1} + \eta_\text{disl}^{-1} \right)^{-1}.
\end{equation}
\end{enumerate}

As an extension of \cite{vankeken2008} we consider cases 1c and 2b with a
curved slab geometry. This curve is described by the parabola
\begin{equation}
y_\text{slab} = -Z \left(\frac{x}{Z}\right)^{n_\text{slab}},
\label{eq:slab_curve}
\end{equation}
where $n_\text{slab} = \num{2.5}$. The inlet velocity is prescribed separately
for the two slab geometry schemes. In the case of the straight subducting slab
$u_\text{inlet} = (\cos(\alpha), -\sin(\alpha))^\top$. And in the case of the
curved slab $u_\text{inlet} = -n$.

\subsubsection{Solution technique}

Note that the nonlinearity in this problem is numerically challenging to
resolve. We first employ a fixed point iterative method to obtain an initial
guess for subsequent solution by Newton's iterative method. Additionally, we
must resolve the pressure discontinuity across the subducting slab. Although
it is suboptimal, we choose the $W^{h,0}$ velocity--pressure--temperature
finite element space where $\mathfrak{s}=1$ as defined in
\cref{sec:problem_definition}, i.e. piecewise constant pressure approximation.

\subsubsection{Results}

\begin{figure}
\centering
\begin{subfigure}{.45\textwidth}
  \centering
  \includegraphics[width=1.\linewidth]{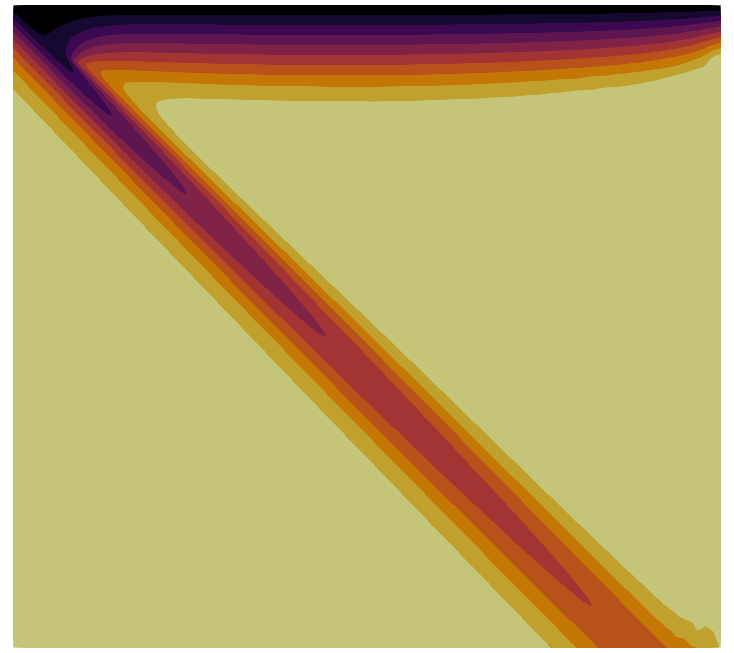} \\
  (a) Case 1c, $u = \tau_r$ on $\Gamma_\text{slab}$
\end{subfigure}%
\begin{subfigure}{.45\textwidth}
  \centering
  \includegraphics[width=1.\linewidth]{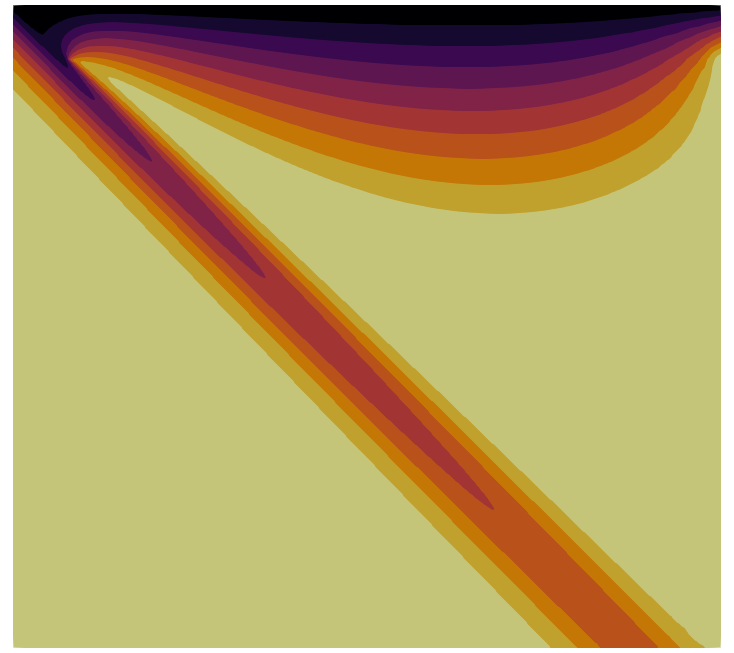} \\
  (b) Case 2b, $u = \tau_r$ on $\Gamma_\text{slab}$
\end{subfigure}
\begin{subfigure}{.45\textwidth}
  \centering
  \includegraphics[width=1.\linewidth]{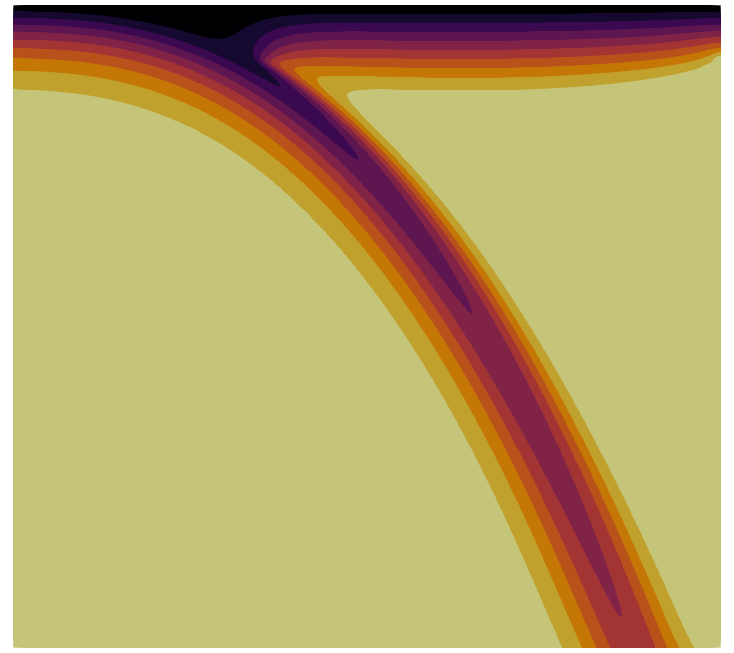} \\
  (c) Case 1c, $u = \tau_r$ on $\Gamma_\text{slab}$
\end{subfigure}%
\begin{subfigure}{.45\textwidth}
  \centering
  \includegraphics[width=1.\linewidth]{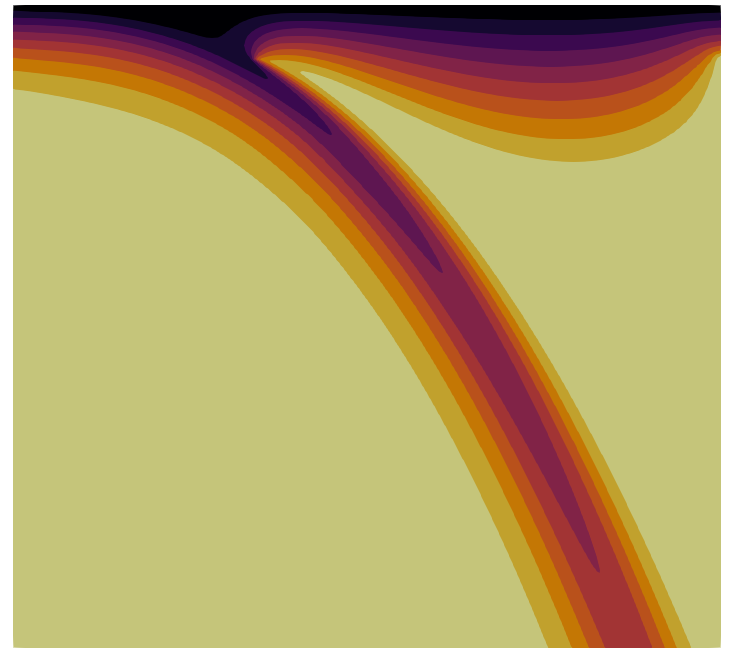} \\
  (d) Case 2b, $u = \tau_r$ on $\Gamma_\text{slab}$
\end{subfigure}
\begin{subfigure}{.45\textwidth}
  \centering
  \includegraphics[width=1.\linewidth]{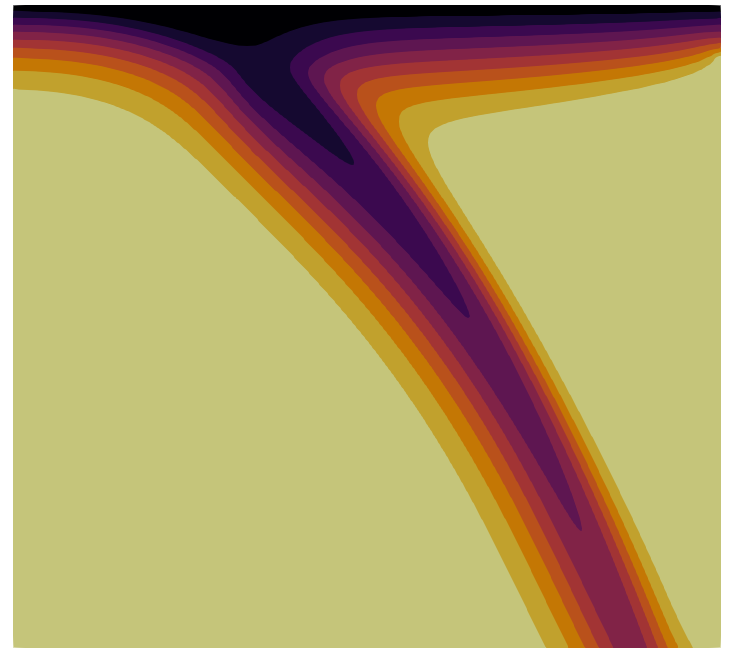} \\
  (e) Case 1c, $u \cdot n = 0$ on $\Gamma_\text{slab}$
\end{subfigure}%
\begin{subfigure}{.45\textwidth}
  \centering
  \includegraphics[width=1.\linewidth]{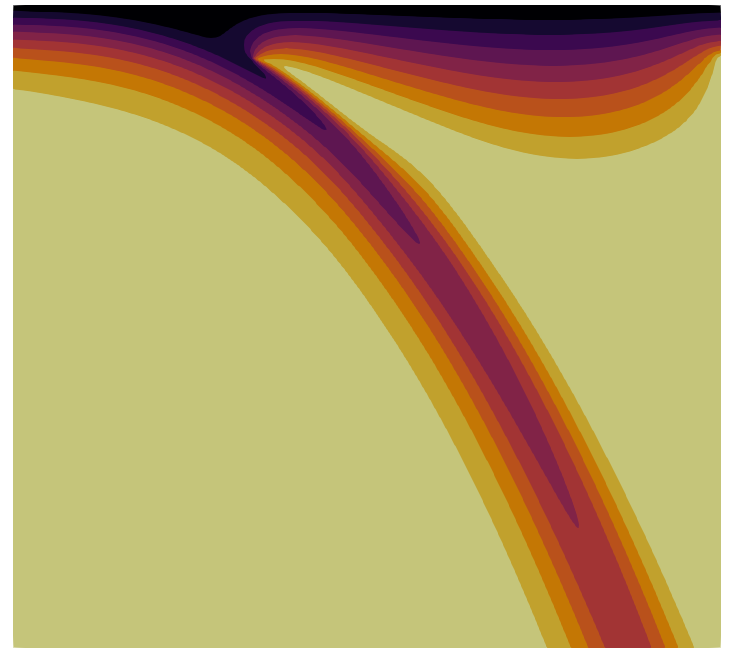} \\
  (f) Case 2b, $u \cdot n = 0$ on $\Gamma_\text{slab}$
\end{subfigure}
\begin{subfigure}{.6\textwidth}
  \centering
  \includegraphics[width=1.\linewidth]{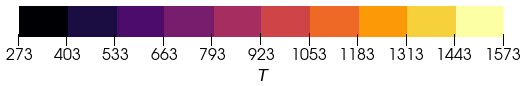}
\end{subfigure}
\caption{Example 4: Subduction zone with: (a) \& (b) straight subducting slab 
where $\alpha = \frac{\pi}{4}$; (c), (d), (e) \& (f) curved subducting slab
defined by \cref{eq:slab_curve} where $n_\text{slab} = 2.5$. The geometry and
model are described in~\cref{sec:subduction_zone}.}
\label{fig:subduction}
\end{figure}

The temperature field from these computations is shown in
\cref{fig:subduction}. In the straight subducting slab computation where $u =
\tau_r$ on $\Gamma_\text{slab}$ the following functionals are computed for
validation and comparison with the community benchmark:
\begin{align}
T_{11,11} &= T(\xi_{11,11}) - T_s, \\
\Vert T_\text{slab} \Vert &= \sqrt{\frac{\sum_{i=1}^{36} \del{T(\xi_{ii}) - T_s}^2}{36}}, \\
\Vert T_\text{wedge} \Vert &= \sqrt{\frac{\sum_{i=10}^{21} \sum_{j=10}^{i} \del{T(\xi_{ij}) - T_s}^2}{78}},
\end{align}
where $\xi_{ij} = (6(i-1), -6(j-1))^\top$ forms a series of discrete equidistant
points with interval spacing of \num{6} originating at $(0, 0)$, see~\cite{vankeken2008}
for details. The computed functionals are tabulated in \cref{tab:subduction_straight}.

\begin{table}[h!]
\centering
\begin{tabular}{llrrr}
Case & Code & $T_{11,11}$      & $\Vert T_\text{slab} \Vert$ & $\Vert T_\text{wedge} \Vert$ \\ \hline
1c   & This work & \num{387.86}     & \num{503.18}            & \num{853.04}        \\
     & Reference & \num{387.84}     & \num{503.13}            & \num{852.92}        \\
2a   & This work & \num{579.43}     & \num{606.50}            & \num{1002.52}       \\
     & Reference & \num{580.66}     & \num{607.11}            & \num{1003.20}       \\
2b   & This work & \num{582.01}     & \num{604.46}            & \num{999.12}        \\
     & Reference & \num{583.36}     & \num{605.11}            & \num{1000.01}       
\end{tabular}
\caption{Example 4: Computed benchmark functionals for comparison
with~\cite{vankeken2008} where the reference data is quoted from the UM
contributing finite element code. The temperature fields computed in these
problems are shown in \cref{fig:subduction}. The minimum and maximum cell
circumradii used in the mesh are \num{0.161} and \num{24.9}, respectively.}
\label{tab:subduction_straight}
\end{table}

\begin{table}[h!]
\centering
\begin{tabular}{llrrrr}
Case & $\Gamma_\text{slab}$ condition & $T_{11,11}$      & $\Vert T_\text{slab} \Vert$ & $\Vert T_\text{wedge} \Vert$ & $T(x_\text{slab}, -60)$  \\ \hline
1c   & $u = \tau_r$ & \num{991.12}     & \num{1093.15}            & \num{1055.66}        & \num{352.80} \\
     & $u \cdot n = 0$ & \num{982.49}     & \num{1066.46}            & \num{1028.21}        &  \num{171.339} \\
2b   & $u = \tau_r$ & \num{961.74}     & \num{1071.79}            & \num{1026.13}        & \num{557.84} \\
     & $u \cdot n = 0$ & \num{961.74}     & \num{1071.73}            & \num{1021.58}       & \num{614.83} \\
\end{tabular}
\caption{Example 4: Computed benchmark functionals in the curved subducting
slab case. Here $n_\text{slab} = \num{2.5}$. The temperature fields computed
in these problems are shown in \cref{fig:subduction}. The minimum and maximum
cell circumradii used in the mesh are \num{0.134} and \num{24.1}, respectively.} 
\label{tab:subduction_curved}
\end{table}

Examine the two cases where $u = \tau_r$ on $\Gamma_\text{slab}$ and the slab
geometry is curved in \cref{fig:subduction}. These cases are the extension of
cases 1c and 2b exhibited in~\cite{vankeken2008}. We introduce one more
functional for with the curved geometry experiments, $T(x_\text{slab},
-60)$. This is the analogy of $T_{11,11}$ for the curved geometry, i.e.,
the temperature measured on the slab--wedge boundary at a depth of
$y_\text{slab} = \num{60}$ where
\begin{equation}
x_\text{slab} = Z \sqrt[\leftroot{-2}\uproot{2}n_\text{slab}]{\frac{-y_\text{slab}}{Z}}.
\end{equation}
The computed functionals from these experiments are shown in
\cref{tab:subduction_curved}.

\begin{figure}
\centering
\begin{subfigure}{.45\textwidth}
  \centering
  \includegraphics[width=1.\linewidth]{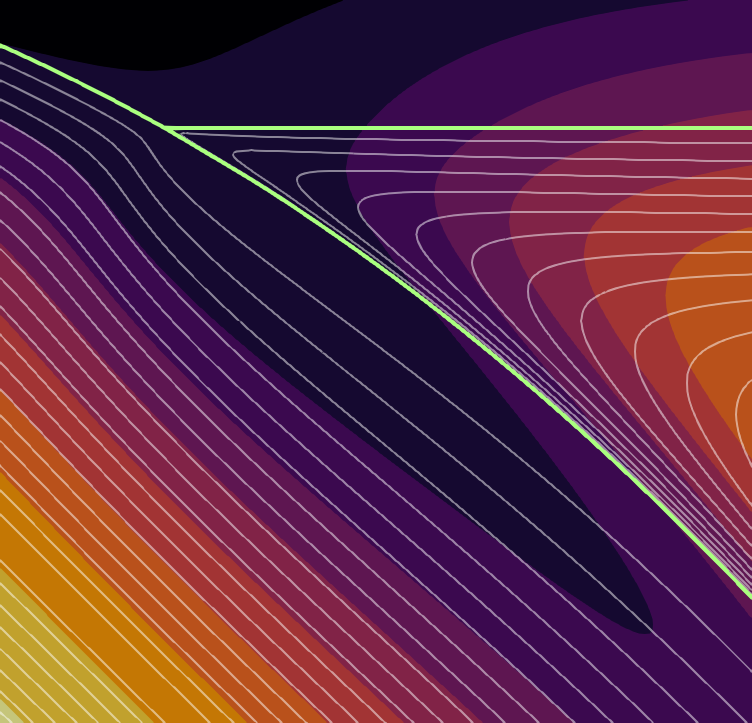} \\
  (a) Case 1c, $u \cdot n = 0$ on $\Gamma_\text{slab}$
\end{subfigure}%
\hspace{0.1cm}
\begin{subfigure}{.45\textwidth}
  \centering
  \includegraphics[width=1.\linewidth]{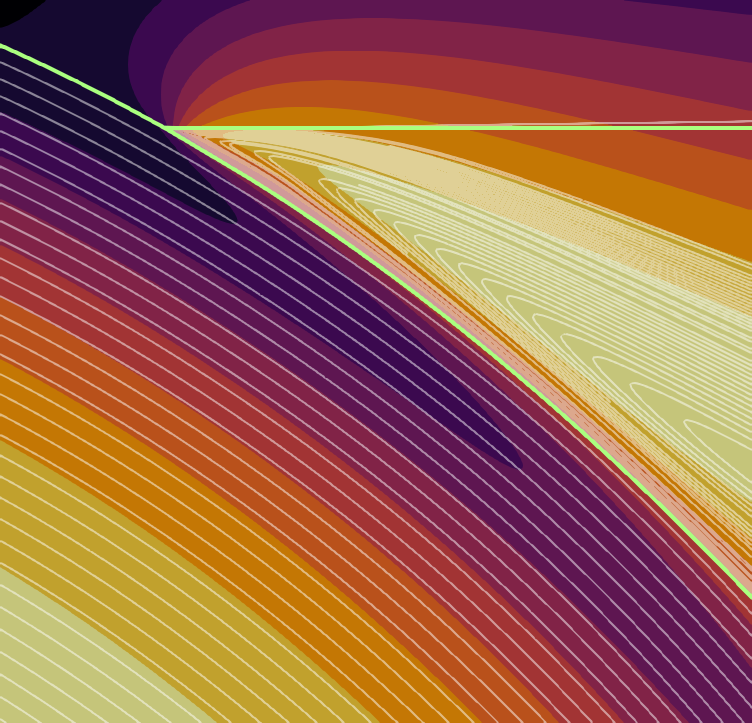} \\
  (b) Case 2b, $u \cdot n = 0$ on $\Gamma_\text{slab}$
\end{subfigure}
\begin{subfigure}{.6\textwidth}
  \centering
  \includegraphics[width=1.\linewidth]{experiments/subduction_benchmark/colourbar.png}
\end{subfigure}
\caption{Example 4: Examination of the streamlines in Cases 1c and 2b where $u
\cdot n = 0$ on $\Gamma_\text{slab}$. The geometry of the subduction interface and
overriding plate is overlaid in green. The temperature fields in the whole
geometry are shown in \cref{fig:subduction}. (a)~shows the overlap of the
no--slip boundary condition on the overriding plate manifests prominently.
(b)~shows that the dislocation creep viscosity model is less affected by this
numerical artifact. The nature of this boundary condition is discussed in
detail in~\cite{vankeken2008}. Here, the streamlines in both (a) and (b) are
generated from \num{200} equidistant source points located on the left ($x =
0$) and right ($x = Z (\tan \alpha)^{-1}) + \delta_x$) boundary.}
\label{fig:subduction_zoom_streamline}
\end{figure}

The plots of the temperature fields from the final two cases where $u \cdot n
= 0$ on $\Gamma_\text{slab}$ and the slab geometry is curved invites
curiosity. There is an apparent change in the mantle flow profile at the plate
depth compared with the previous cases. These changes are primarily due to the
benchmark~\cite{vankeken2008} requiring the overlap of the discontinuity $u =
0$ on the plate boundary and $u = \tau_r$ on the upper subducting slab. The
lower subducting slab on $\Gamma_\text{slab}$ and boundary condition $u \cdot
n = 0$ propagates the single node data on which $u = 0$ is enforced. This
results in the velocity `bump' observed. Examination of the velocity
streamlines elucidates this as shown in \cref{fig:subduction_zoom_streamline}.
The work~\cite{vankeken2008} discusses the issue of this boundary condition
overlap in greater detail. If we were to enforce the free slip boundary
condition strongly, indeed we may isolate the single degree of freedom
associated with the overriding plate and subducting slab node and enforce a
free slip condition. Nitsche's method is mathematically consistent and does
not permit such `variational crimes'.

\subsection{Example 5: Iterative solvers and 3D geometry}
\label{sec:scaling}

In this final example we demonstrate the viability of Nitsche's method with
Krylov subspace iterative solvers. Rigorous demonstration of scalable solvers
for finite element problems in computational geodynamics is available
in~\cite{may2008preconditioned}. In this work we are concerned with the
conditioning of the underlying finite element matrix, subject to the penalty
parameter in Nitsche's method, $C_\text{IP} \ell^2 / h_F$
(see~\cref{sec:sipg_dirichlet,sec:sipg_freeslip}). With the experiments
exhibited in this section we seek to alleviate the worry that including a
penalty parameter in the finite element formulation will cause preconditioned
Krylov subspace solvers to fail to converge.

Motivated by the results in the previous section, we consider the case 1c,
where $u \cdot n = 0$ on $\Gamma_\text{slab}$. We further choose the curved
subducting slab geometry where $n_\text{slab} = 1.5$. Our numerical experiment
in this section will compute the finite element approximation  on a
hierarchical sequence of meshes. The first mesh is a coarse representation of
the geometry, and the sequence defines progressively finer meshes which are
adaptively generated based on residual--based error estimation of the finite
element solution. Other examples of adaptive refinement using the tools
showcased in this work are demonstrated in~\cite{nate2018}.

We will solve the problem in a 2D and 3D geometry. The 2D geometry is as
described in the previous section. To generate the 3D geometry we extrude the
2D geometry by rotating $\frac{\pi}{4}$ radians about the $(0, 1, 0)$ axis
centred at $(-Z_\text{plate}, 0, 0)$. On the new near and far side faces
we apply the free slip boundary condition. All other boundary conditions are
the extrusion of the 2D case.

Newton's iterative method is used to solve the finite element system. We solve
the underlying linear system using an outer fGMRES solver with an inverse
viscosity pressure mass matrix preconditioner for the pressure block,
cf.~\cite{may2008preconditioned}. We construct the linear solver as follows:
for the velocity block we apply a GMRES smoother with PETSc's geometric
algebraic multigrid preconditioner with smoothed aggregation; on the pressure
block a conjugate gradient smoother with HYPRE-BoomerAMG preconditioner; on
the temperature block we use a GMRES smoother with HYPRE-BoomerAMG
preconditioner.

\begin{figure}
\centering
\begin{subfigure}{.5\textwidth}
  \centering
  \includegraphics[width=1.\linewidth]{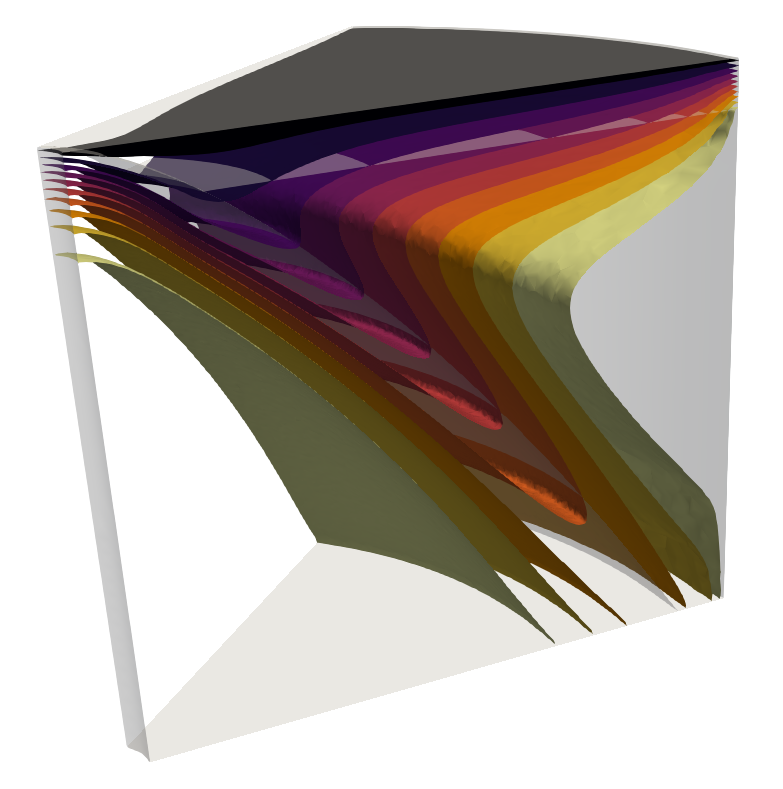}
\end{subfigure}
\begin{subfigure}{.5\textwidth}
  \centering
  \includegraphics[width=1.\linewidth]{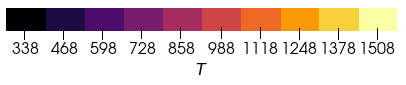} 
\end{subfigure}
\caption{Example 5: Temperature field isosurfaces from the 3D computation
overlaying the semitransparent extruded geometry described in \cref{sec:scaling}. Here
$n_\text{slab} = 1.5$.}
\label{fig:result_3d}
\end{figure}

In each experiment we measure the total number of degrees of freedom (DoF),
the maximum and minimum mesh cell dihedral angles $h_\text{max}$ and
$h_\text{min}$, respectively, the number of fGMRES iterations
$n_\text{Krylov}$, the number of Newton iterations $n_\text{Newton}$ and the
final finite element vector residual in the 2--norm $\Vert r \Vert_2$. The
tabulated results of the two experiments are shown in
\cref{tab:iterative_method_its}. The computed finite element temperature field
in the 3D geometry is shown in \cref{fig:result_3d}.

\begin{table}[h!]
\centering
\begin{tabular}{rrS[table-format=2.2]S[table-format=3.2]rrr}
& DoF & $h_\text{max}$ & $h_\text{min}$ & $n_\text{Krylov}$ & $n_\text{Newton}$ & $\Vert r \Vert_2$ \\ \hline
2D & \num{14956} & 28.6 & 13.5 & \num{45} & \num{4} & \num{7.7545e-09} \\
& \num{34136} & 25.4 & 6.51 & \num{45} & \num{4} & \num{8.2716e-09} \\
& \num{77226} & 25.7 & 3.45 & \num{47} & \num{4} & \num{9.2493e-09} \\
& \num{168552} & 25.5 & 1.82 & \num{45} & \num{4} & \num{5.0943e-09} \\
& \num{358146} & 25.2 & 0.889 & \num{44} & \num{4} & \num{7.6659e-09} \\
& \num{747281} & 25.9 & 0.481 & \num{43} & \num{4} & \num{7.5715e-09} \\
& \num{1542503} & 24.0 & 0.220 & \num{43} & \num{4} & \num{4.7911e-09} \\
& \num{3156039} & 25.2 & 0.111 & \num{42} & \num{4} & \num{9.6823e-09} \\
& \num{6338077} & 68.0 & 0.0342 & \num{42} & \num{4} & \num{5.9607e-09} \\
& \num{14248941} & 46.9 & 0.0227 & \num{39} & \num{4} & \num{8.0802e-09} \\ 
& \num{20979279} & 46.7 & 0.0114 & \num{56} & \num{5} & \num{4.9678e-09} \\ 
\\\hline
3D & \num{324556} & 50.7 & 16.2 & \num{154} & \num{5} & \num{6.5248e-09} \\
& \num{1012848} & 46.1 & 8.71 & \num{108} & \num{5} & \num{8.5585e-09} \\
& \num{3656383} & 45.1 & 4.10 & \num{176} & \num{5} & \num{9.4709e-09} \\
& \num{5495723} & 42.6 & 3.11 & \num{156} & \num{5} & \num{9.3598e-09}
\end{tabular}
\caption{Example 5: Examination of the viability of free slip boundary
conditions enforced via Nitsche's method when using a Krylov subspace
iterative solver. Viability is indicated by a near--constant number of Krylov
subspace iterations required to find convergence of the finite element
residual. Here, we consider satisfactory convergence when $\Vert r \Vert_2 <
\num{e-8}.$}
\label{tab:iterative_method_its}
\end{table}






\section{Conclusion}
\label{sec:conclusion}

The formulation and examples presented in this work elucidate the use of
Nitsche's method for complex problems in geodynamics which require free slip
boundary conditions. Additionally we have presented a flexible computational
implementation which requires only a symbolic representation of the numerical
flux function $\mathcal{F}^v(\cdot,\cdot)$ to automatically formulate Nitsche
boundary conditions. The author emphasises that we are not limited to the
Stokes equations nor free slip boundary conditions with the developed
computational tools.

\subsubsection*{Acknowledgements}

The authors wish to thank P.~E.~van~Keken of the Carnegie Institution for
Science, Department of Terrestrial Magnetism for his advice, particularly
regarding~\cref{sec:subduction_zone}. NS gratefully acknowledges the support
of Carnegie Institution for Science President's Fellowship.

\appendix
\section{Blankenbach et al. and Tosi et al. benchmark results}
\label{sec:blankenbach_tosi_tables}

In this section we tabulate the results computed from a reproduction of the
Blankenbach~et~al.~\cite{blankenbach1989} and Tosi~et~al.~\cite{tosi2015}
benchmarks. This data is shown in \cref{tab:blankenbach_results_full,tab:tosi_results_full},
respectively. The functionals of interest computed from the benchmark FEM
solutions are shown in \cref{tab:benchmark_functionals}.
%


\begin{table}[h!]
\centering
\begin{tabular}{ll}
    Blankenbach~et~al.~\cite{blankenbach1989} & \\ \hline
    Nusselt number  &
        $\mathrm{Nu} = 
        \frac{\int_{\Gamma_\text{top}} \nabla T \cdot n \dif{s}}
        {\int_{\Gamma_\text{bottom}} T \dif{s}}$ \\
    Heat flux & $\xi_i = \left.\left(\nabla T \cdot n\right)\right|_{x=x_i}$ \\ 
        & $x_1 = (0, 0), x_2 = (L, 0),$ \\
        & $x_3 = (0, H), x_4 = (L, H)$ \\
    Tosi~et~al.~\cite{tosi2015} & \\ \hline
    Top Nusselt number &
    $\mathrm{Nu}_\text{top} = 
    \frac{\int_{\Gamma_\text{top}} \nabla T \cdot n \dif{s}}
    {\int_{\Gamma_\text{bottom}} T \dif{s}}$ \\
    Bottom Nusselt number &
    $\mathrm{Nu}_\text{bottom} = 
    \frac{\int_{\Gamma_\text{bottom}} \nabla T \cdot n \dif{s}}
    {\int_{\Gamma_\text{top}} T \dif{s}}$ \\
    Average temperature &
    $\langle T \rangle = \frac{\int_\Omega T \dif{x}}{\int_\Omega \dif{x}}$ \\
    Root-mean-square (RMS) velocity &
    $u_\text{rms} = \sqrt{\frac{\int_\Omega u \cdot u \dif{x}}{\int_\Omega \dif{x}}}$ \\
    Surface RMS velocity & 
    $u_\text{surf rms} = \sqrt{\frac{\int_{\Gamma_\text{top}} u \cdot u \dif{s}}{\int_{\Gamma_\text{top}} \dif{s}}}$ \\
    Average rate of viscous dissipation & 
    $\langle \Phi \rangle = \frac{\int_\Omega 2 \eta \varepsilon(u) : \varepsilon(u) \dif{x}}{\int_\Omega \dif{x}}$ \\
    Average work done against gravity & $\langle W \rangle = \frac{\int_\Omega T u_y \dif{x}}{\int_\Omega \dif{x}}$
\end{tabular}
\caption{The functionals computed in the Blankenbach~et~al. and Tosi~et~al. benchmarks. Here
$\Gamma_\text{top}$ and $\Gamma_\text{bottom}$ correspond to the exterior boundaries at
$y = H$ and $y = 0$, respectively.}
\label{tab:benchmark_functionals}
\end{table}

\begin{table}
\centering

\begin{subtable}{\textwidth}
\centering
\makeblankentable{\blankstrongone}
\caption{Case 1: strong}
\end{subtable}
\begin{subtable}{\textwidth}
\centering
\makeblankentable{\blankweakone}
\caption{Case 1: weak}
\end{subtable}

\begin{subtable}{\textwidth}
\centering
\makeblankentable{\blankstrongtwo}
\caption{Case 2: strong}
\end{subtable}
\begin{subtable}{\textwidth}
\centering
\makeblankentable{\blankweaktwo}
\caption{Case 2: weak}
\end{subtable}

\begin{subtable}{\textwidth}
\centering
\makeblankentable{\blankstrongthree}
\caption{Case 3: strong}
\end{subtable}
\begin{subtable}{\textwidth}
\centering
\makeblankentable{\blankweakthree}
\caption{Case 3: weak}
\end{subtable}

\begin{subtable}{\textwidth}
\centering
\makeblankentable{\blankstrongfour}
\caption{Case 4: strong}
\end{subtable}
\begin{subtable}{\textwidth}
\centering
\makeblankentable{\blankweakfour}
\caption{Case 4: weak}
\end{subtable}

\begin{subtable}{\textwidth}
\centering
\makeblankentable{\blankstrongfive}
\caption{Case 5: strong}
\end{subtable}
\begin{subtable}{\textwidth}
\centering
\makeblankentable{\blankweakfive}
\caption{Case 5: weak}
\end{subtable}

\caption{Blankenbach~et~al. benchmark results. We provide a comparison between strong
and weak imposition of the free slip boundary condition. The functionals computed
in this experiment match well with the work of Blankenbach~et~al.~\cite{blankenbach1989}}
\label{tab:blankenbach_results_full}
\end{table}


\begin{table}
\centering

\begin{subtable}{\textwidth}
\centering
\maketositable{\tosistrongone}
\caption{Case 1: strong}
\end{subtable}

\begin{subtable}{\textwidth}
\centering
\maketositable{\tosiweakone}
\caption{Case 1: weak}
\end{subtable}

\begin{subtable}{\textwidth}
\centering
\maketositable{\tosistrongtwo}
\caption{Case 2: strong}
\end{subtable}

\begin{subtable}{\textwidth}
\centering
\maketositable{\tosiweaktwo}
\caption{Case 2: weak}
\end{subtable}

\begin{subtable}{\textwidth}
\centering
\maketositable{\tosistrongthree}
\caption{Case 3: strong}
\end{subtable}

\begin{subtable}{\textwidth}
\centering
\maketositable{\tosiweakthree}
\caption{Case 3: weak}
\end{subtable}

\begin{subtable}{\textwidth}
\centering
\maketositable{\tosistrongfour}
\caption{Case 4: strong}
\end{subtable}

\begin{subtable}{\textwidth}
\centering
\maketositable{\tosiweakfour}
\caption{Case 4: weak}
\end{subtable}

\caption{Tosi~et~al. benchmark results. A comparison of the imposition of
the free slip boundary condition in a strong and weak sense is provided.
The results computed here match well with the work compiled by
Tosi~et~al.~\cite{tosi2015}}
\label{tab:tosi_results_full}
\end{table}

\bibliographystyle{abbrvnat}
\bibliography{references}
\end{document}